\documentclass[%
aip,
amsmath,amssymb,
reprint,%
]{revtex4-1}

\usepackage{hyperref}
\usepackage[usenames]{color}
\usepackage[pdftex]{graphicx}
\usepackage{amsmath, amstext, amssymb, amsfonts, amsxtra}
\usepackage{textcomp}
\usepackage{xspace}
\DeclareSymbolFontAlphabet{\amsmathbb}{AMSb} 
\usepackage{soul}
\usepackage{xcolor}
\usepackage{epstopdf}
\usepackage{braket}
\usepackage{float}
\usepackage{enumitem}
\setlist{nosep}

\newcommand{\rhop}{\hat{\rho}} 
\newcommand{\sop}{\hat{\sigma}} 
\newcommand{\Vop}{\hat{V}}

\newcommand{\Lop}{\mathcal{L}}
\newcommand{\IPR}{\mathcal{I}} 
\newcommand{\AIPR}{\bar{\mathcal{I}}}    
\newcommand{\CIPR}{\mathcal{I}_{\rm C}} 
\newcommand{\ACIPR}{\bar{\mathcal{I}}_{\rm C}}

\newcommand{\tr}{{\rm tr}}

\newcommand{\Hop}{\hat{H}}

\newcommand{\Dop}{\mathcal{D}}

\newcommand{\im}{{\rm i}}

\newcommand{\smt}{Science, Mathematics and Technology Cluster, Singapore University of Technology and Design, 8 Somapah Road, 487372 Singapore} 
\newcommand{\epd}{EPD Pillar, Singapore University of Technology and Design, 8 Somapah Road, 487372 Singapore}

\begin{document}

\title{Localization-delocalization effects of a delocalizing dissipation on disordered XXZ spin chains}     

\author{Xiansong Xu} 
\affiliation{\smt}
\author{Dario Poletti}
\email{dario\_poletti@sutd.edu.sg}
\affiliation{\smt} 
\affiliation{\epd}

\begin{abstract}
The interplay between interaction, disorder, and dissipation has shown a rich phenomenology. Here we investigate a disordered XXZ spin chain in contact with a bath which, alone, would drive the system towards a highly delocalized and coherent Dicke state. We show that there exist regimes for which the natural orbitals of the single-particle density matrix of the steady state are all localized in the presence of strong disorders, either for weak interaction or strong interaction. We show that the averaged steady-state occupation in the eigenbasis of the open system Hamiltonian could follow an exponential decay for intermediate disorder strength in the presence of weak interactions, while it is more evenly spread for strong disorder or for stronger interactions. 
Last, we show that strong dissipation increases the coherence of the steady states, thus reducing the signatures of localization. We capture such signatures of localization also with a concatenated inverse participation ratio which simultaneously takes into account how localized are the eigenstates of the Hamiltonian, and how close is the steady state to an incoherent mixture of different energy eigenstates.      
\end{abstract}

\maketitle

\begin{quotation}
As discovered in 1958 by P. W. Anderson \cite{Anderson1958}, in the presence of disorder, quantum systems can become insulators, or in other words the systems become localized.
When the particles in the quantum system interact with each other, the effects of disorders can be very different from the non-interacting case, known as the many-body localization \cite{BaskoAltshuler2006}, and the study of this topic is still attracting significant attention.
In recent years, researchers have also studied how the environment surrounding such disordered and interacting systems can affect the localization properties of quantum systems. It was found that depending on the magnitude of disorder and on the environment, one can still observe possible signatures of localization in these quantum systems.
Here we study the effects of a particular type of environment which tries to drive the system towards a highly delocalized state, while the presence of disorder, together with the interaction between the particles, could result in a localized system. We find that even when the system shows signatures of localization, there can be interesting differences in its long-time properties. 
\end{quotation}

\section{Introduction} \label{sec: intro}

The effects of disorder in quantum systems took a prominent place in 1958 with the discovery of Anderson localization \cite{Anderson1958} in non-interacting systems. The characterization of the effects of disorder in interacting systems occurred much later \cite{BaskoAltshuler2006, OganesyanHuse2007, PalHuse2010, GiamarchiSchulz1987, GiamarchiSchulz1988} with the study of many-body localized (MBL) systems. Some reviews on this phase of matter which does not thermalize are Refs. \onlinecite{NandkishoreHuse2015, AbaninPapic2017, AbaninSerbyn2019}.    
MBL systems have also been studied experimentally with ultracold atoms \cite{SchreiberBloch2015, SchreiberBloch2015a, BordiaSchneider2016, ChoiGross2016, BordiaBloch2017, BordiaBloch2017a, LuschenSchneider2017}, trapped ions \cite{SmithMonroe2016} and superconducting qubits \cite{RoushanMartinis2017, GuoWang2020}.     

A natural question to ask is how MBL systems respond when coupled to a bath \cite{NandkishoreHuse2014, JohriBhatt2015, Nandkishore2015, ZnidaricVarma2016, MedvedyevaZnidaric2016, LeviGarrahan2016, FischerAltman2016, LuschenSchneider2017, HyattBauer2017, EverestLevi2017, LuitzDeRoeck2017, NandkishoreGopalakrishnan2017, ZnidaricGoold2017, ZnidaricLjubotina2018, KarlssonVerdozzi2018, VakulchykDenisov2018, XuPoletti2018, WuEckardt2019a, Rubio-AbadalGross2019, WyboPollmann2020a}. Studies of strongly interacting and strongly disordered systems coupled to small systems that act as baths have shown that the overall system can become localized or thermalized \cite{Nandkishore2015, HyattBauer2017, LuitzDeRoeck2017}. 
Analyzing the spectral features of an MBL system coupled to a bath has shown that there can exist regimes, e.g., with large enough interaction and weak enough coupling to the baths, such that the spectral response to a local spin flip is gapped, an indication of localization \cite{NandkishoreHuse2014, JohriBhatt2015}. This can occur despite other quantities, such as the level statistics, do not show signatures of localization for couplings to the baths which decrease exponentially with the system size.  
Coupling MBL systems to local dissipative boundary driving has shown that in the ergodic phase one can observe both diffusive and subdiffusive transport \cite{ZnidaricVarma2016, ZnidaricGoold2017, ZnidaricLjubotina2018}. 
Even when an interacting and disordered system is coupled to a bath that drives it to the infinite temperature state, the relaxation dynamics can be significantly slowed down, even possibly assuming a stretched exponential form \cite{MedvedyevaZnidaric2016, LeviGarrahan2016, FischerAltman2016, EverestLevi2017}.    
It is thus clear that strongly disordered and interacting systems can show a large variety of phenomenology depending on the system and bath parameters. 

Recently in Refs. \onlinecite{VakulchykDenisov2018, XuPoletti2018}, the authors studied a disordered XXZ spin chain coupled to a bath which, alone, would tend to drive the system towards a highly delocalized and coherent Dicke state. Such dissipator has been first proposed in \onlinecite{DiehlZoller2008}, in which it was shown how it could be realized in ultracold atoms experiments, and later on it was implemented with a trapped-ion universal simulator \cite{SchindlerBlatt2013}. In Refs. \onlinecite{VakulchykDenisov2018, XuPoletti2018} it was shown that despite the presence of such delocalizing dissipator, for large enough disorder and interactions, it is possible to find signatures compatible with the localization of the steady state. Notably, the level spacing of the density matrix follows a Poisson distribution, and the natural orbitals of the single-particle density matrix can be all exponentially localized in steady states. 

In this work we go more in-depth in this investigation. 
First we show that for a small disorder the density matrix can show a very different amount of off-diagonal terms in the spin configuration basis depending on the magnitudes of disorder and interaction. 
In particular, for the isotropic case of the XXZ chain and weak disorder, the steady state approaches a Dicke state. For large disorders and/or large interactions, the steady state density matrix is, instead, mostly diagonal. 
For strong disorder, we find two regions of the parameter space in which the natural orbitals of the single-particle density matrix are localized, either with weak or with strong interaction. 
We also analyze the steady-state density matrix in the basis of the eigenstates of the system Hamiltonian and we find that for moderately large disorders the steady-state occupation in eigenenergy basis follows an approximately exponentially decay, indicating that the steady state is mainly composed of few (localized) many-body eigenstates of the Hamiltonian. In the regime of large interactions instead, the steady state has a more evenly spread overlap with the eigenstates of the Hamiltonian, indicating a clear difference from the previous regime. We also show that increasing the dissipation strength for this particular bath increases the coherence in the steady-state density matrix thus significantly weakening the signatures of localization. We quantify this with a concatenated inverse participation ratio which signals whether the eigenstates of the Hamiltonian are localized and, at the same time, the steady state is not a coherent (hence delocalized) mixture of these energy eigenstates.

\section{Model} \label{sec: model}

We consider a dissipative spin chain whose evolution is described by a master equation   
\begin{align}
  \label{eq:master}
  \frac{\partial \rhop}{\partial t}&=\Lop\left[\rhop\right]=-\frac{\im}{\hbar}\left[\Hop,\rhop\right]+\Dop\left[\rhop\right],  
\end{align}
where $\Lop$ is the system's Lindbladian, $\Dop$ the portion of the master equation which includes the effects of dissipation and $\Hop$ the Hamiltonian. We consider a prototypical MBL Hamiltonian given by an XXZ spin-$1/2$ chain with local disorders
\begin{align}
  \label{eq:ham}
  \Hop =\sum_{l=1}^{L-1}\left[J \left(\sop^{x}_l \sop^x_{l+1}+\sop^y_l \sop^y_{l+1}\right)+\Delta \sop^z_l \sop^z_{l+1}\right]+\sum_{l=1}^{L} h_l \sop^z_l, 
\end{align}
where the elements of $\sop_l^\alpha$ are given by the Pauli matrices for $\alpha=x,\;y, \;z$, $J$ is the tunneling amplitude, $\Delta$ is the interaction strength and the random field $h_l$ is uniformly distributed in $\left[-W,W\right]$ with $W$ characterizing the disorder strength. In the following we use open boundary conditions. Throughout this article We consider a system size of $L=10$ and work in the zero total magnetization sector. Note that this Hamiltonian exactly maps to an interacting spinless fermionic chain after Jordan-Wigner transformation \cite{JordanWigner1928, LiebMattis1961}. In the absence of dissipation, this model has been shown to have many-body localized or ergodic phases depending on the strengths of disorder and interaction \cite{PalHuse2010, BeraBardarson2015, BeraBardarson2017, LezamaBardarson2017, LinHeidrich-Meisner2018, LuitzAlet2015, SerbynAbanin2015}, although for a better understanding of the properties of this system in the thermodynamic limit, further studies on the role of finite size effects may be needed \cite{ SerbynAbanin2015, SuntajsVidmar2020a, AbaninVasseur2021, PandaZnidaric2019a, Kiefer-EmmanouilidisSirker2021}. 
We consider a dissipator in Gorini-Kossakowski-Sudarshan-Lindblad (GKSL) form \cite{GoriniSudarshan1976, Lindblad1976} given by       
\begin{align}
  \mathcal{D}\left[\rhop\right]&=\gamma\sum_{l=1}^{L-1} \left(\Vop_{l,l+1}\rhop \Vop_{l,l+1}^\dagger-\frac{1}{2}\left\lbrace \Vop_{l,l+1}^\dagger \Vop^{}_{l,l+1}, \rhop \right\rbrace \right),  \label{eq:diss}  
\end{align}
where $\gamma$ indicates the coupling strength, $\{.,.\}$ indicates the anticommutator, and the jump operators $\Vop^{}_{l,l+1}$ are     
\begin{align}
  \label{eq:jump} 
  \Vop^{}_{l,l+1}=\left(\sop^+_l+\sop^+_{l+1}\right)\left(\sop^-_l-\sop^-_{l+1}\right), 
\end{align} 
where $\sop^+_l=\left(\sop^x_l+\im\sop^y_l\right)/2$ and $\sop^-=\left(\sop_l^x-\im\sop_l^y\right)/2$. 
Such a dissipator has been first proposed in bosonic systems \cite{DiehlZoller2008, DiehlZoller2010, TomadinZoller2011}, and subsequently further studied with bosons and also with fermions and spins \cite{YiZoller2012, BardynDiehl2013}. While in Ref. \onlinecite{DiehlZoller2008} the authors proposed a realization for bosonic systems by immersion of the system in a superfluid bath, an actual experimental realization was performed using an ion-trap universal quantum simulator \cite{SchindlerBlatt2013}. It is important to stress that proving that a disordered open many-body system, with a large number of small energy gaps in the spectrum, can be faithfully described by a Markovian master equation in Eq. (\ref{eq:master}) with dissipator given in Eq. (\ref{eq:diss}) is not a trivial task, however a realization via Trotterization of the evolution operator \cite{Lloyd1996} on a quantum simulator is in any case possible as long as one uses small enough time steps to approximate the exact evolution with the Trotterized one. It is thus possible to study also experimentally the model from Eqs. (\ref{eq:master}-\ref{eq:diss}).     

The interest of analyzing the effects of dissipator given by Eq. (\ref{eq:diss}) in a system with Hamiltonian (\ref{eq:ham}) is that, for large enough disorder, the system has eigenstates with a large and sudden variability of local magnetization, but the dissipator we have chosen $\Vop^{}_{l,l+1}$ favors a balanced magnetization on neighboring sites. 
Hence, the model (\ref{eq:master}-\ref{eq:diss}) can be an important test-bed to analyze possible properties of disordered many-body open quantum systems. 
We stress that this system has been studied for a single particle problem in Refs. \onlinecite{YusipovIvanchenko2017, VershininaLaptyeva2017} where it was shown that for large enough disorder the system showed signatures of localization. Here, as in Refs. \onlinecite{VakulchykDenisov2018, XuPoletti2018}, we add the non-trivial effect of interactions. 
To compute the steady-state $\rhop_s$ we used either exact diagonalization or time evolutions and we concentrated on the sector with zero total magnetization so as to probe the strong effects of the interaction.

\section{Results} \label{sec: results}    

\begin{figure}
  \includegraphics[width=\columnwidth]{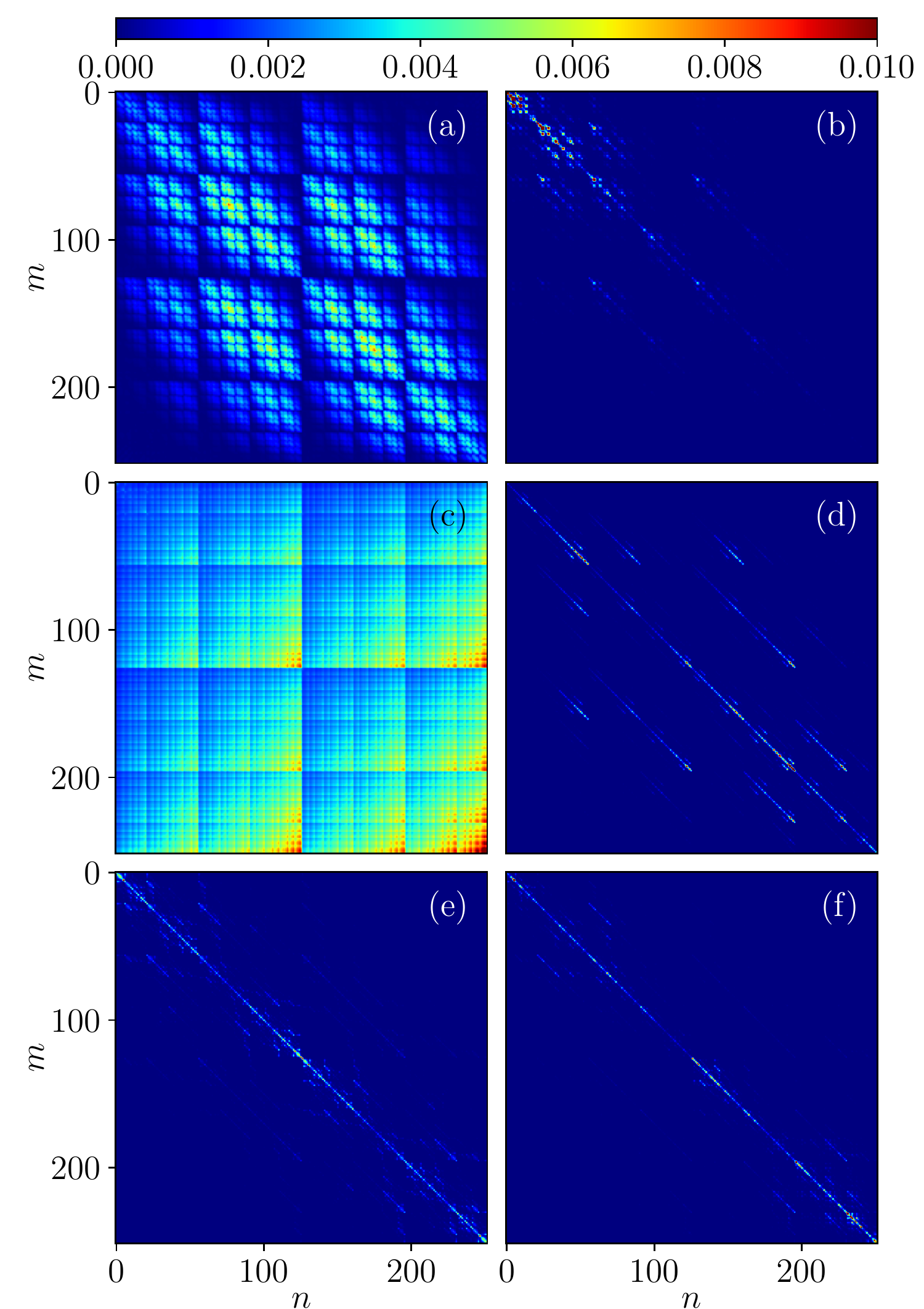}
  \caption{Single realizations for the absolute value of steady-state density matrix in the spin configurations basis with increasing interaction strength (a, b) $\Delta=0.5J$, (c, d) $\Delta=1J$, (e, f) $\Delta=7J$. Left panels (a, c, d) have weak disorder $W=0.1J$ whereas right panels (b, d, f) have disorder strength of $W=12J$. The dissipation strength $\hbar\gamma=0.1J$ and the system size $L=10$. } 
  \label{fig:density_matrix}   
\end{figure} 

\begin{figure}
  \includegraphics[width=\columnwidth]{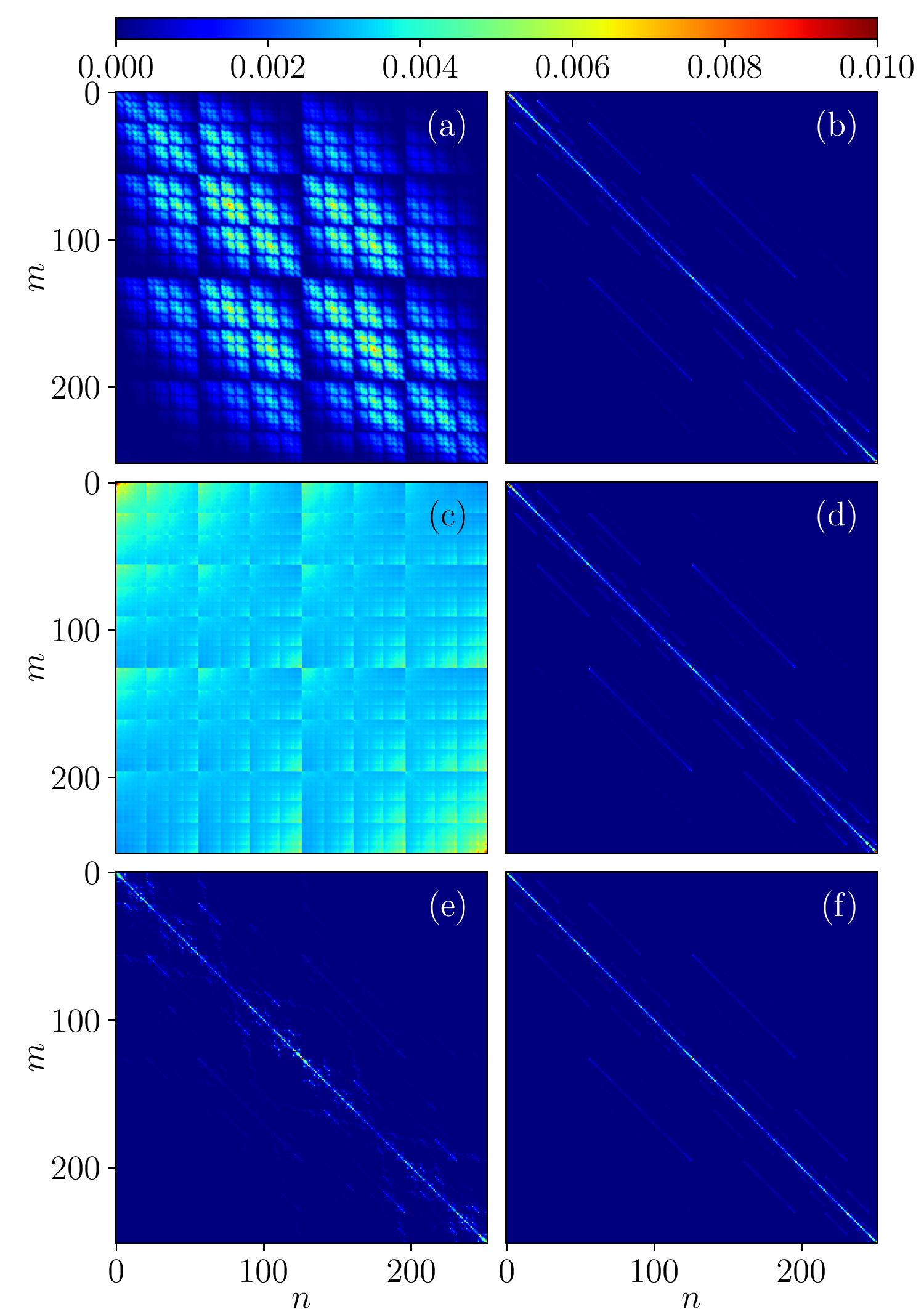}
  \caption{Averaged absolute value of density matrix over 100 disorder realizations in the spin configurations basis with increasing interaction strength (a, b) $\Delta=0.5J$, (c, d) $\Delta=1J$, (e, f) $\Delta=7J$. Left panels (a, c, d) have weak disorder $W=0.1J$ whereas right panels (b, d, f) have disorder strength of $W=12J$.  The dissipation strength $\hbar\gamma=0.1J$ and the system size $L=10$.} 
  \label{fig:density_matrix_average}      
\end{figure}

To begin, we would like to give the reader an overview of the possible steady state properties of the system for different magnitudes of disorder and interaction. In Fig. \ref{fig:density_matrix} we plot the absolute values of the elements of the density matrix in the basis formed by all possible spin configurations. In these plots we represent the density matrix for a randomly chosen disorder distribution, and the results are typical. In the left panels (a), (c), and (e) we consider a small disorder strength $W=0.5J$, while in the right columns we have significantly larger disorder $W=12J$. The different rows represent possible interaction values: $\Delta=0.5J$ for panels (a) and (b), $\Delta=J$ for panels (c) and (d), and $\Delta=7J$ for panels (e) and (f). For weak interaction and weak disorder, Fig. \ref{fig:density_matrix}(a), one observes large coherence, however, the steady state becomes even more striking for weak disorder and $\Delta=J$ as the state approximates the Dicke state $|\psi_D\rangle$ (symmetric superposition of all the possible spin configurations) which is the exact steady state in the absence of disorder \cite{XuPoletti2018}. For larger interaction or larger disorder, Figs. \ref{fig:density_matrix}(b, d, e, f), the off-diagonal elements are significantly reduced and the steady state is much more diagonal. However we can observe that for a given disorder realization the density matrix appears to have different structures in presence of large disorder and weak interaction, Fig. \ref{fig:density_matrix}(b), or vice versa, Fig. \ref{fig:density_matrix}(e), or even both large interaction and disorder, Fig. \ref{fig:density_matrix}(f).          
In Figs. \ref{fig:density_matrix_average} we perform a parallel of Fig. \ref{fig:density_matrix} where we first average the density matrix over 100 disorder realizations before plotting the absolute value of its elements. From Figs. \ref{fig:density_matrix_average} we can deduce that for weak disorders the density matrices of different realizations are very similar to each other, as if the steady state is not strongly affected by the presence of disorder. We also observe that, on average, the density matrices are mostly diagonal for other cases.

One way used to characterize isolated disordered many-body systems is via the single-particle density matrix $\rho_{\mathrm{sp}}^{j,k}=\langle \psi_n | \sop^+_j \sop^-_k | \psi_n \rangle$ for each eigenstate $|\psi_n\rangle$ of the Hamiltonian \cite{BeraBardarson2015, BeraBardarson2017, LezamaBardarson2017, LinHeidrich-Meisner2018}. In particular, it was shown that for large enough disorder with either weak or strong interaction, the natural orbitals of the single-particle density matrix, i.e., $\Psi_{\alpha}$ such that $\rho_{\mathrm{sp}}\Psi_{\alpha}=n_{\alpha}\Psi_{\alpha}$ where $n_{\alpha}$ is the normalized occupation spectrum, all become exponentially localized. 
As done in Ref. \onlinecite{XuPoletti2018}, one can characterize in the same manner the steady state of an open and disordered many-body quantum system from the single-particle density matrix $\rho_{\mathrm{sp}}^{j,k}=\tr(\sop^+_j \sop^-_k \rhop_{\rm s})$. 
We then compute a weighted inverse participation ratio 
\begin{align}
  \IPR(\hat{\rho}_{\rm s}) =\sum_{\alpha,l} n_{\alpha}|\Psi_{\alpha}(l)|^4 \label{eq:ipr}
\end{align}
and its average over the disorder realizations $\AIPR$.   

\begin{figure}
  \includegraphics[width=\columnwidth]{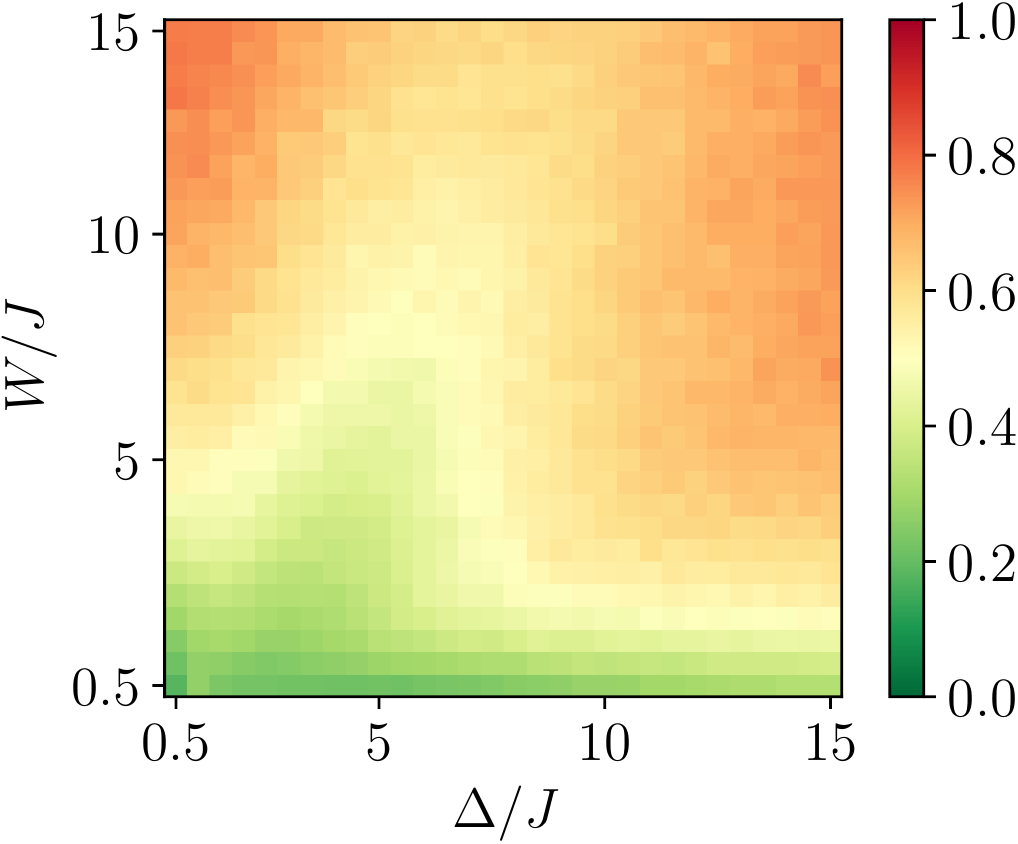}
  \caption{Color map of averaged weighted inverse participation ratio $\AIPR$ as a function of disorder strength $W$ and interaction strength $\Delta$. Parameters are $L=10$, $\hbar\gamma=0.1J$, and $100$ disorder realizations. }
  \label{fig:IPR} 
\end{figure}

In Fig. \ref{fig:IPR} we show a color map of the averaged IPR, $\AIPR$, versus the interaction $\Delta$ and disorder strength $W$. For weak disorders we observe that $\AIPR$ is small indicating the presence of delocalized natural orbitals in the steady state. For intermediate values of interaction, around $\Delta/J\approx 4$ we observe that such delocalized orbitals are relevant in the steady state even for larger values of disorders. For large disorder $W$ we observe that either weak or strong interaction $\Delta$ results in large IPR.

\begin{figure}
  \includegraphics[width=\columnwidth]{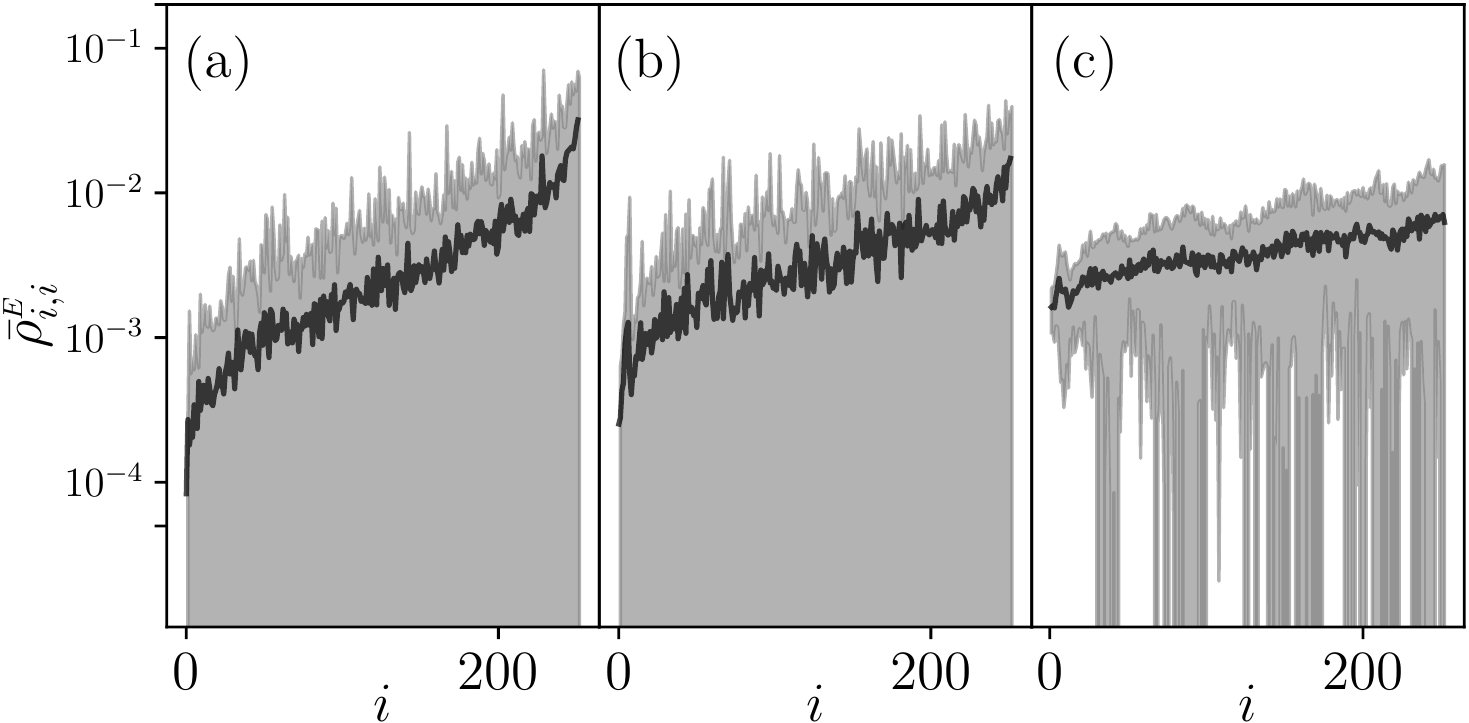}
  \caption{Averaged diagonal elements of density matrix $\bar{\rho}^{_E}_{i,i}$ over 100 disorder realizations versus the energy level $i$ for increasing value of the energy. In (a) $\Delta=0.5J$, $W=7J$, (b) $\Delta=0.5J$, $W=12J$, and (c) $\Delta=15J$, $W=7J$. The grey shaded region demonstrates the standard deviation from the averaged results. The values of $\AIPR$ for the cases shown in the different panels are (a) $\AIPR = 0.60 \pm 0.01$, (b) $\AIPR = 0.73\pm 0.01$ and (c) $\AIPR = 0.713\pm 0.008$.} 
  \label{fig:project_H}   
\end{figure}     

At this point we try to characterize whether the steady states of the system have similar properties in the regions with large $\AIPR$, whether for small or large interaction $\Delta$ in the presence of strong disorder. To do so we first define the steady state density matrix in the energy basis $\rho^{_E}_{i,j}$ as 
\begin{align}
  \rhop_s = \sum_{i,j} \rho^{_E}_{i,j} |\psi_i\rangle\langle \psi_j|
\end{align} 
where $|\psi_i\rangle$ is the $i$-th eigenstates of the Hamiltonian. 
In Fig. \ref{fig:project_H} we then show the diagonal terms of $\rho^{_E}_{i,i}$ in ascending order of the eigenenergies. The black curve represents the average over 100 disorder realizations $\bar{\rho}^{_E}_{i,i}$ while the gray shading is the standard deviation, which follows closely the functional form of the average values. We observe that for $W=7J$ and $\Delta=0.5J$, Fig. \ref{fig:project_H}(a), the steady state has an exponentially decreasing overlap with the eigenstates of the Hamiltonian. This indicates that few (localized) states dominate the steady state. For the Hamiltonian chosen and the particular type of non-thermal dissipator we are studying, the most occupied states tend to be those with higher energy. 
For larger disorder $W=12J$ and $\Delta=0.5J$, Fig. \ref{fig:project_H}(b), the exponential decay is less pronounced. 
In Fig. \ref{fig:project_H}(c), for $W=7J$ and $\Delta=15J$ the slope of the profile has an overall small tilt due to interactions. This indicates that the steady state is closer to an even mixture of energy eigenstates. 
Hence, while both for weak and large interaction $\AIPR$ is large, the steady state can have significantly different properties. 

\begin{figure}
  \includegraphics[width=\columnwidth]{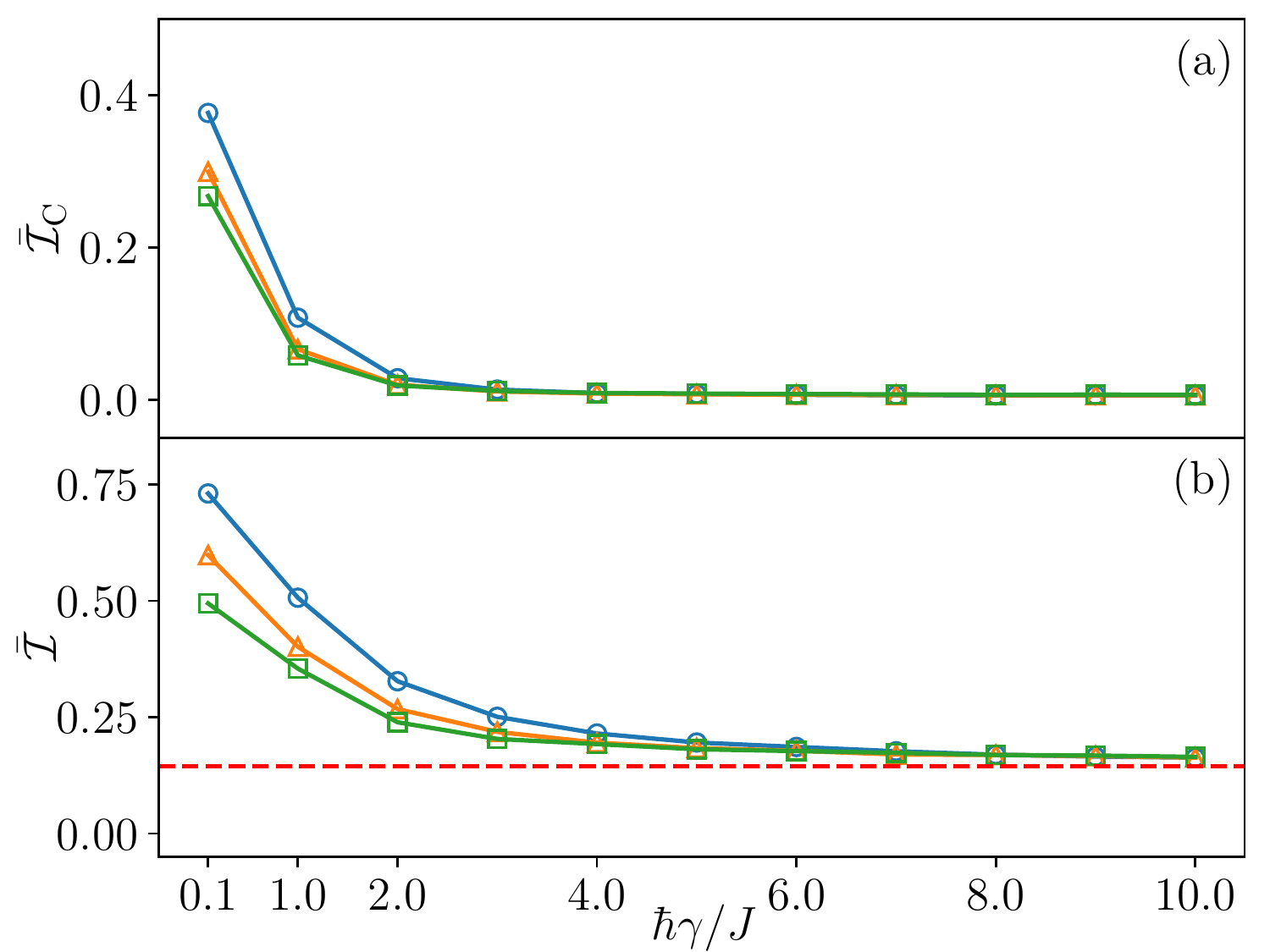}
  \caption{(a) Averaged concatenated inverse participation ratio $\bar{\mathcal{I}}_{\rm C}$ as a function of the dissipation strength $\gamma$. (b) Averaged inverse participation ratio for $\bar{\mathcal{I}}$ as a function of the dissipation strength $\gamma$ for various disorder strength $W=12J$ (cyan circles), $W=7J$ (orange triangles), $W=5J$ (green squares) with a common interaction strength $\Delta=0.5J$. The red dashed line marks the finite size value of $\IPR$ for a Dicke state. Other parameters are $L=10$ and $100$ disorder realizations. } 
  \label{fig:CIPR}   
\end{figure}

When a disordered system is coupled to a bath, the localization properties could be due to the fact the Hamiltonian is localized or that the dissipator dominates the dynamics and drives the system towards a steady state which is localized, or a combination of the two effects. To be able to characterize whether the localization properties of a steady state are more prominently inherited from the Hamiltonian, we introduce the concatenated inverse participation ratio $\CIPR$ which measures whether a steady state is, at the same time, diagonal in the energy basis of the Hamiltonian and the energy eigenstates are localized. More precisely, we define $\CIPR$ as    
\begin{align}
  \CIPR(\rhop_s)=\sum_{i,j} N_i|\langle \psi_j|\varphi_i\rangle|^4  \mathcal{I}(\Ket{\psi_j}\Bra{\psi_j}) \label{eq:cipr}
\end{align}   
where $\mathcal{I}(\Ket{\psi_j}\Bra{\psi_j})$ is the inverse participation ratio of the natural orbitals of the $j$-th eigenstate of the Hamiltonian $|\psi_j\rangle$ given in Eq. (\ref{eq:ipr}) \cite{BeraBardarson2015}. In Eq. (\ref{eq:cipr}), $|\varphi_i\rangle$ is the $i$-th eigenstates of the steady-state density operator, $\rhop_s = \sum_i N_i|\varphi_i \rangle\langle \varphi_i|$. In Eq. (\ref{eq:cipr}) $|\langle \psi_j|\varphi_i\rangle|^4$ can give a large contribution only when $|\varphi_i\rangle$ has significant overlap with only a single $|\psi_j\rangle$ \footnote{We assume no degeneracies}, and $\mathcal{I}(\Ket{\psi_j}\Bra{\psi_j})$ is large if that eigenstate is localized. Hence, $\CIPR$ can indeed inform us on whether the steady state is made of an incoherent mixture of localized energy eigenstates.      
In Fig. \ref{fig:CIPR} we show both the averaged $\CIPR$ from Eq. (\ref{eq:cipr}), $\ACIPR$ in panel (a), and the averaged $\IPR$ from Eq. (\ref{eq:ipr}), $\AIPR$ in panel (b), both averaged over $100$ disorder realizations. In both panels, we showed the inverse participation ratios for different disorder strengths versus the magnitude of the dissipation rate $\gamma$. We observe that for small $\gamma$, and large enough disorder, the $\CIPR$ is large and it decreases as $\gamma$ increases. 
From comparing $\CIPR$ with $\AIPR$, we observe that they have a similar trend, indicating that the localization properties that we observe in the steady state are, in a sense, inherited from the Hamiltonian.     
In the limit of large $\gamma$ both inverse participation ratios obtain a minimum. We note that $\AIPR$ tends towards a finite value which may seem counter-intuitive given the delocalizing nature of the dissipator. This can be explained by the fact that even for the Dicke state (which is the state that the dissipator alone would produce), $\IPR(|\psi_D\rangle\langle\psi_D|)$ reaches a small yet finite value which we indicate, in Fig. \ref{fig:CIPR}(b) by the red dashed line. This asymptotic value would decrease as the system size increases.

\begin{figure}
  \includegraphics[width=\columnwidth]{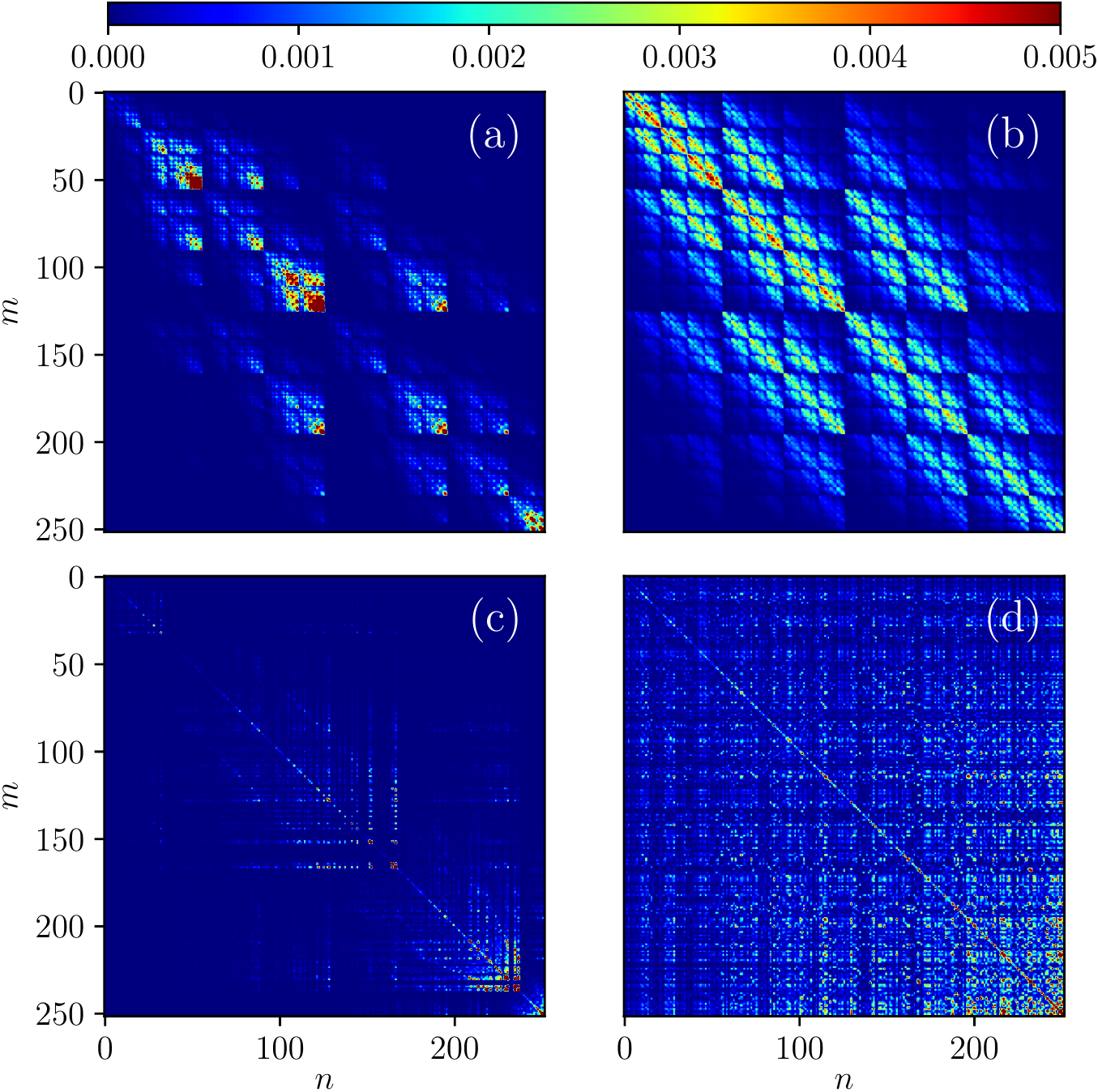}
  \caption{Single realizations of density matrix in spin configurations basis (a, b) and in eigenbasis of the system (c, d) an intermediate dissipation strength $\hbar\gamma=1J$ in left panels (a, c) and large dissipation strength  $\hbar\gamma=10J$ in right panels (c, d). Common parameters are $\Delta=0.5J$, $W=12J$ and $L=10$.} 
  \label{fig:dissipation}   
\end{figure}

The fact that in this model the dissipator tends to delocalize the system can also be observed in Fig. \ref{fig:dissipation} in which we show the modulus of the elements of the density matrix both in the spin configuration, panels (a, b), and in the energy basis, panels (c, d). In panels (a, c) we considered an intermediate value of the dissipation rate $\hbar\gamma=J$, and in panels (b, d) we chose a much larger value  $\hbar\gamma=10J$. The latter case shows a significantly increased coherence in the steady-state density matrix.

\section{Conclusions} \label{sec: conclusions}

We have studied a disordered interacting spin chain coupled to an external bath. While disorder and interaction can lead the system to be localized, the bath tends to induce homogeneous states for the system. 
The rich variety of attainable steady states in this system is exemplified in Figs. (\ref{fig:density_matrix}, \ref{fig:density_matrix_average}), in which we can observe that for small disorder there is large coherence even far from the diagonal, and for $\Delta=J$ the steady state well approximates the Dicke state. Larger values of disorder or interaction make the density matrix close to diagonal both in the spin configuration basis and in the energy eigenbasis of the system, especially after averaging out over many disorder realizations.  
It is thus an interesting set-up to study disordered many-body open quantum systems, as already shown in Refs. \onlinecite{VakulchykDenisov2018, XuPoletti2018}. 

While Refs. \onlinecite{VakulchykDenisov2018, XuPoletti2018} focus on possible signatures of many-body localization in open quantum systems, this article focuses on the localization-delocalization properties within the open quantum systems and gives an attempt to measure the localization for steady state directly with respect to the system Hamiltonian. This is done by first extending the studies of IPR by probing the phase diagram of the system in steady states. We show that the averaged inverse participation ratio of the natural orbitals of the single-particle density matrix can be large when the disorder is large enough, both for weak and strong interactions. 

We then revisit the meaning of the localization in many-body open quantum systems by analyzing the steady-state density matrix in the Hamiltonian eigenbasis and we find that for intermediate-large disorders the steady state is composed of an exponentially decaying distribution of energy eigenstates, each of which is localized. However, as disorder or interaction increases, the distribution of eigenstates flattens out and more eigenstates have a significant overlap with the steady state. 
This informs us that while the inverse participation ratio can be very large, the steady state can be very different, and only in regimes with intermediate-large disorders, and weak interaction, we obtain a steady state in which a few energy eigenstates are more prominent.  

Last, we proposed the concatenated inverse participation ratio which takes into account both the ``diagonality'' of the density matrix in energy eigenbasis of the system Hamiltonian as well as the localization properties of the basis states. This allows us to probe whether the signatures of localization are intrinsic of the system Hamiltonian or more from an interplay of Hamiltonian and dissipative terms. For the model studied here we observe that when the inverse participation ratio is large, the steady state is also close to diagonal in the energy basis, indicating that its localization properties are inherited from the Hamiltonian. Increasing the dissipation rates increases the coherence and decreases the localization properties of the steady state. 
In the future, larger systems sizes should be studied, and other types of non-thermal baths, so as to learn more about possible phases of matter emerging from the interplay of interaction, disorder, and tailored dissipators. It would also be interesting to study the relaxation dynamics of two-time correlations and out-of-time-order correlations for many-body localized systems \cite{FanZhai2017} with dissipation.

\begin{acknowledgments}
D.P. acknowledges support from the Ministry of Education of Singapore AcRF MOE Tier-II (Project No. MOE2018-T2-2-142) and fruitful discussions with D. Abanin, B. Gr\'{e}maud, F. Heidrich-Meisner, and R. Chen. Part of the computational work was performed on resources of the National Supercomputing Centre, Singapore \cite{NSCC}.
\end{acknowledgments}

\section*{Data availability}
The data that support the findings of this study are available from the corresponding author upon reasonable request.


\begin{thebibliography}{64}%
  \makeatletter
  \providecommand \@ifxundefined [1]{%
    \@ifx{#1\undefined}
  }%
  \providecommand \@ifnum [1]{%
    \ifnum #1\expandafter \@firstoftwo
  \else \expandafter \@secondoftwo
\fi
}%
\providecommand \@ifx [1]{%
  \ifx #1\expandafter \@firstoftwo
\else \expandafter \@secondoftwo
 \fi
}%
\providecommand \natexlab [1]{#1}%
\providecommand \enquote  [1]{``#1''}%
\providecommand \bibnamefont  [1]{#1}%
\providecommand \bibfnamefont [1]{#1}%
\providecommand \citenamefont [1]{#1}%
\providecommand \href@noop [0]{\@secondoftwo}%
\providecommand \href [0]{\begingroup \@sanitize@url \@href}%
\providecommand \@href[1]{\@@startlink{#1}\@@href}%
\providecommand \@@href[1]{\endgroup#1\@@endlink}%
\providecommand \@sanitize@url [0]{\catcode `\\12\catcode `\$12\catcode
`\&12\catcode `\#12\catcode `\^12\catcode `\_12\catcode `\%12\relax}%
\providecommand \@@startlink[1]{}%
\providecommand \@@endlink[0]{}%
\providecommand \url  [0]{\begingroup\@sanitize@url \@url }%
\providecommand \@url [1]{\endgroup\@href {#1}{\urlprefix }}%
\providecommand \urlprefix  [0]{URL }%
\providecommand \Eprint [0]{\href }%
\providecommand \doibase [0]{http://dx.doi.org/}%
\providecommand \selectlanguage [0]{\@gobble}%
\providecommand \bibinfo  [0]{\@secondoftwo}%
\providecommand \bibfield  [0]{\@secondoftwo}%
\providecommand \translation [1]{[#1]}%
\providecommand \BibitemOpen [0]{}%
\providecommand \bibitemStop [0]{}%
\providecommand \bibitemNoStop [0]{.\EOS\space}%
\providecommand \EOS [0]{\spacefactor3000\relax}%
\providecommand \BibitemShut  [1]{\csname bibitem#1\endcsname}%
\let\auto@bib@innerbib\@empty
\bibitem [{\citenamefont {Anderson}(1958)}]{Anderson1958}%
\BibitemOpen
\bibfield  {author} {\bibinfo {author} {\bibfnamefont {P.~W.}\ \bibnamefont
    {Anderson}},\ }\bibfield  {title} {\enquote {\bibinfo {title} {Absence of
  {{Diffusion}} in {{Certain Random Lattices}}},}\ }\href {\doibase
  10.1103/PhysRev.109.1492} {\bibfield  {journal} {\bibinfo  {journal} {Phys.
  Rev.}\ }\textbf {\bibinfo {volume} {109}},\ \bibinfo {pages} {1492--1505}
(\bibinfo {year} {1958})}\BibitemShut {NoStop}%
\bibitem [{\citenamefont {Basko}, \citenamefont {Aleiner},\ and\ \citenamefont
{Altshuler}(2006)}]{BaskoAltshuler2006}%
\BibitemOpen
\bibfield  {author} {\bibinfo {author} {\bibfnamefont {D.}~\bibnamefont
  {Basko}}, \bibinfo {author} {\bibfnamefont {I.}~\bibnamefont {Aleiner}}, \
  and\ \bibinfo {author} {\bibfnamefont {B.}~\bibnamefont {Altshuler}},\
  }\bibfield  {title} {\enquote {\bibinfo {title} {Metal\textendash insulator
      transition in a weakly interacting many-electron system with localized
single-particle states},}\ }\href {\doibase 10.1016/j.aop.2005.11.014}
{\bibfield  {journal} {\bibinfo  {journal} {Ann. Phys. (N.Y.)}\ }\textbf
  {\bibinfo {volume} {321}},\ \bibinfo {pages} {1126--1205} (\bibinfo {year}
{2006})}\BibitemShut {NoStop}%
\bibitem [{\citenamefont {Oganesyan}\ and\ \citenamefont
{Huse}(2007)}]{OganesyanHuse2007}%
\BibitemOpen
\bibfield  {author} {\bibinfo {author} {\bibfnamefont {V.}~\bibnamefont
    {Oganesyan}}\ and\ \bibinfo {author} {\bibfnamefont {D.~A.}\ \bibnamefont
    {Huse}},\ }\bibfield  {title} {\enquote {\bibinfo {title} {Localization of
  interacting fermions at high temperature},}\ }\href {\doibase
  10.1103/PhysRevB.75.155111} {\bibfield  {journal} {\bibinfo  {journal} {Phys.
  Rev. B}\ }\textbf {\bibinfo {volume} {75}},\ \bibinfo {pages} {155111}
(\bibinfo {year} {2007})}\BibitemShut {NoStop}%
\bibitem [{\citenamefont {Pal}\ and\ \citenamefont {Huse}(2010)}]{PalHuse2010}%
\BibitemOpen
\bibfield  {author} {\bibinfo {author} {\bibfnamefont {A.}~\bibnamefont
  {Pal}}\ and\ \bibinfo {author} {\bibfnamefont {D.~A.}\ \bibnamefont {Huse}},\
  }\bibfield  {title} {\enquote {\bibinfo {title} {Many-body localization phase
  transition},}\ }\href {\doibase 10.1103/PhysRevB.82.174411} {\bibfield
  {journal} {\bibinfo  {journal} {Phys. Rev. B}\ }\textbf {\bibinfo {volume}
{82}},\ \bibinfo {pages} {174411} (\bibinfo {year} {2010})}\BibitemShut
{NoStop}%
\bibitem [{\citenamefont {Giamarchi}\ and\ \citenamefont
{Schulz}(1987)}]{GiamarchiSchulz1987}%
\BibitemOpen
\bibfield  {author} {\bibinfo {author} {\bibfnamefont {T.}~\bibnamefont
    {Giamarchi}}\ and\ \bibinfo {author} {\bibfnamefont {H.~J.}\ \bibnamefont
    {Schulz}},\ }\bibfield  {title} {\enquote {\bibinfo {title} {Localization and
{{Interaction}} in {{One}}-{{Dimensional Quantum Fluids}}},}\ }\href
{\doibase 10.1209/0295-5075/3/12/007} {\bibfield  {journal} {\bibinfo
  {journal} {Europhys. Lett.}\ }\textbf {\bibinfo {volume} {3}},\ \bibinfo
{pages} {1287--1293} (\bibinfo {year} {1987})}\BibitemShut {NoStop}%
\bibitem [{\citenamefont {Giamarchi}\ and\ \citenamefont
{Schulz}(1988)}]{GiamarchiSchulz1988}%
\BibitemOpen
\bibfield  {author} {\bibinfo {author} {\bibfnamefont {T.}~\bibnamefont
    {Giamarchi}}\ and\ \bibinfo {author} {\bibfnamefont {H.~J.}\ \bibnamefont
    {Schulz}},\ }\bibfield  {title} {\enquote {\bibinfo {title} {Anderson
  localization and interactions in one-dimensional metals},}\ }\href {\doibase
  10.1103/PhysRevB.37.325} {\bibfield  {journal} {\bibinfo  {journal} {Phys.
  Rev. B}\ }\textbf {\bibinfo {volume} {37}},\ \bibinfo {pages} {325--340}
(\bibinfo {year} {1988})}\BibitemShut {NoStop}%
\bibitem [{\citenamefont {Nandkishore}\ and\ \citenamefont
{Huse}(2015)}]{NandkishoreHuse2015}%
\BibitemOpen
\bibfield  {author} {\bibinfo {author} {\bibfnamefont {R.}~\bibnamefont
    {Nandkishore}}\ and\ \bibinfo {author} {\bibfnamefont {D.~A.}\ \bibnamefont
    {Huse}},\ }\bibfield  {title} {\enquote {\bibinfo {title} {Many-{{Body
          Localization}} and {{Thermalization}} in {{Quantum Statistical
Mechanics}}},}\ }\href {\doibase 10.1146/annurev-conmatphys-031214-014726}
{\bibfield  {journal} {\bibinfo  {journal} {Annu. Rev. Condens. Matter
  Phys.}\ }\textbf {\bibinfo {volume} {6}},\ \bibinfo {pages} {15--38}
(\bibinfo {year} {2015})}\BibitemShut {NoStop}%
\bibitem [{\citenamefont {Abanin}\ and\ \citenamefont
{Papi{\'c}}(2017)}]{AbaninPapic2017}%
\BibitemOpen
\bibfield  {author} {\bibinfo {author} {\bibfnamefont {D.~A.}\ \bibnamefont
    {Abanin}}\ and\ \bibinfo {author} {\bibfnamefont {Z.}~\bibnamefont
    {Papi{\'c}}},\ }\bibfield  {title} {\enquote {\bibinfo {title} {Recent
      progress in many-body localization: {{Recent}} progress in many-body
  localization},}\ }\href {\doibase 10.1002/andp.201700169} {\bibfield
  {journal} {\bibinfo  {journal} {Ann. Phys. (Berlin)}\ }\textbf {\bibinfo
  {volume} {529}},\ \bibinfo {pages} {1700169} (\bibinfo {year}
{2017})}\BibitemShut {NoStop}%
\bibitem [{\citenamefont {Abanin}\ \emph {et~al.}(2019)\citenamefont {Abanin},
  \citenamefont {Altman}, \citenamefont {Bloch},\ and\ \citenamefont
{Serbyn}}]{AbaninSerbyn2019}%
\BibitemOpen
\bibfield  {author} {\bibinfo {author} {\bibfnamefont {D.~A.}\ \bibnamefont
  {Abanin}}, \bibinfo {author} {\bibfnamefont {E.}~\bibnamefont {Altman}},
  \bibinfo {author} {\bibfnamefont {I.}~\bibnamefont {Bloch}}, \ and\ \bibinfo
{author} {\bibfnamefont {M.}~\bibnamefont {Serbyn}},\ }\bibfield  {title}
{\enquote {\bibinfo {title} {Colloquium: {{Many}}-body localization,
  thermalization, and entanglement},}\ }\href {\doibase
  10.1103/RevModPhys.91.021001} {\bibfield  {journal} {\bibinfo  {journal}
  {Rev. Mod. Phys.}\ }\textbf {\bibinfo {volume} {91}},\ \bibinfo {pages}
{021001} (\bibinfo {year} {2019})}\BibitemShut {NoStop}%
\bibitem [{\citenamefont {Schreiber}\ \emph
  {et~al.}(2015{\natexlab{a}})\citenamefont {Schreiber}, \citenamefont
  {Hodgman}, \citenamefont {Bordia}, \citenamefont {L{\"u}schen}, \citenamefont
  {Fischer}, \citenamefont {Vosk}, \citenamefont {Altman}, \citenamefont
{Schneider},\ and\ \citenamefont {Bloch}}]{SchreiberBloch2015}%
\BibitemOpen
\bibfield  {author} {\bibinfo {author} {\bibfnamefont {M.}~\bibnamefont
    {Schreiber}}, \bibinfo {author} {\bibfnamefont {S.~S.}\ \bibnamefont
  {Hodgman}}, \bibinfo {author} {\bibfnamefont {P.}~\bibnamefont {Bordia}},
  \bibinfo {author} {\bibfnamefont {H.~P.}\ \bibnamefont {L{\"u}schen}},
  \bibinfo {author} {\bibfnamefont {M.~H.}\ \bibnamefont {Fischer}}, \bibinfo
  {author} {\bibfnamefont {R.}~\bibnamefont {Vosk}}, \bibinfo {author}
  {\bibfnamefont {E.}~\bibnamefont {Altman}}, \bibinfo {author} {\bibfnamefont
    {U.}~\bibnamefont {Schneider}}, \ and\ \bibinfo {author} {\bibfnamefont
    {I.}~\bibnamefont {Bloch}},\ }\bibfield  {title} {\enquote {\bibinfo {title}
    {Observation of many-body localization of interacting fermions in a
quasirandom optical lattice},}\ }\href {\doibase 10.1126/science.aaa7432}
{\bibfield  {journal} {\bibinfo  {journal} {Science}\ }\textbf {\bibinfo
  {volume} {349}},\ \bibinfo {pages} {842--845} (\bibinfo {year}
{2015}{\natexlab{a}})}\BibitemShut {NoStop}%
\bibitem [{\citenamefont {Schreiber}\ \emph
  {et~al.}(2015{\natexlab{b}})\citenamefont {Schreiber}, \citenamefont
  {Hodgman}, \citenamefont {Bordia}, \citenamefont {L{\"u}schen}, \citenamefont
  {Fischer}, \citenamefont {Vosk}, \citenamefont {Altman}, \citenamefont
{Schneider},\ and\ \citenamefont {Bloch}}]{SchreiberBloch2015a}%
\BibitemOpen
\bibfield  {author} {\bibinfo {author} {\bibfnamefont {M.}~\bibnamefont
    {Schreiber}}, \bibinfo {author} {\bibfnamefont {S.~S.}\ \bibnamefont
  {Hodgman}}, \bibinfo {author} {\bibfnamefont {P.}~\bibnamefont {Bordia}},
  \bibinfo {author} {\bibfnamefont {H.~P.}\ \bibnamefont {L{\"u}schen}},
  \bibinfo {author} {\bibfnamefont {M.~H.}\ \bibnamefont {Fischer}}, \bibinfo
  {author} {\bibfnamefont {R.}~\bibnamefont {Vosk}}, \bibinfo {author}
  {\bibfnamefont {E.}~\bibnamefont {Altman}}, \bibinfo {author} {\bibfnamefont
    {U.}~\bibnamefont {Schneider}}, \ and\ \bibinfo {author} {\bibfnamefont
    {I.}~\bibnamefont {Bloch}},\ }\bibfield  {title} {\enquote {\bibinfo {title}
    {Observation of many-body localization of interacting fermions in a
quasirandom optical lattice},}\ }\href {\doibase 10.1126/science.aaa7432}
{\bibfield  {journal} {\bibinfo  {journal} {Science}\ }\textbf {\bibinfo
  {volume} {349}},\ \bibinfo {pages} {842--845} (\bibinfo {year}
{2015}{\natexlab{b}})}\BibitemShut {NoStop}%
\bibitem [{\citenamefont {Bordia}\ \emph {et~al.}(2016)\citenamefont {Bordia},
  \citenamefont {L{\"u}schen}, \citenamefont {Hodgman}, \citenamefont
  {Schreiber}, \citenamefont {Bloch},\ and\ \citenamefont
{Schneider}}]{BordiaSchneider2016}%
\BibitemOpen
\bibfield  {author} {\bibinfo {author} {\bibfnamefont {P.}~\bibnamefont
    {Bordia}}, \bibinfo {author} {\bibfnamefont {H.~P.}\ \bibnamefont
    {L{\"u}schen}}, \bibinfo {author} {\bibfnamefont {S.~S.}\ \bibnamefont
  {Hodgman}}, \bibinfo {author} {\bibfnamefont {M.}~\bibnamefont {Schreiber}},
  \bibinfo {author} {\bibfnamefont {I.}~\bibnamefont {Bloch}}, \ and\ \bibinfo
{author} {\bibfnamefont {U.}~\bibnamefont {Schneider}},\ }\bibfield  {title}
{\enquote {\bibinfo {title} {Coupling {{Identical}} one-dimensional
  {{Many}}-{{Body Localized Systems}}},}\ }\href {\doibase
  10.1103/PhysRevLett.116.140401} {\bibfield  {journal} {\bibinfo  {journal}
  {Phys. Rev. Lett.}\ }\textbf {\bibinfo {volume} {116}},\ \bibinfo {pages}
{140401} (\bibinfo {year} {2016})}\BibitemShut {NoStop}%
\bibitem [{\citenamefont {Choi}\ \emph {et~al.}(2016)\citenamefont {Choi},
  \citenamefont {Hild}, \citenamefont {Zeiher}, \citenamefont {Schau{\ss}},
  \citenamefont {{Rubio-Abadal}}, \citenamefont {Yefsah}, \citenamefont
  {Khemani}, \citenamefont {Huse}, \citenamefont {Bloch},\ and\ \citenamefont
{Gross}}]{ChoiGross2016}%
\BibitemOpen
\bibfield  {author} {\bibinfo {author} {\bibfnamefont {J.-y.}\ \bibnamefont
  {Choi}}, \bibinfo {author} {\bibfnamefont {S.}~\bibnamefont {Hild}}, \bibinfo
  {author} {\bibfnamefont {J.}~\bibnamefont {Zeiher}}, \bibinfo {author}
  {\bibfnamefont {P.}~\bibnamefont {Schau{\ss}}}, \bibinfo {author}
  {\bibfnamefont {A.}~\bibnamefont {{Rubio-Abadal}}}, \bibinfo {author}
  {\bibfnamefont {T.}~\bibnamefont {Yefsah}}, \bibinfo {author} {\bibfnamefont
    {V.}~\bibnamefont {Khemani}}, \bibinfo {author} {\bibfnamefont {D.~A.}\
    \bibnamefont {Huse}}, \bibinfo {author} {\bibfnamefont {I.}~\bibnamefont
    {Bloch}}, \ and\ \bibinfo {author} {\bibfnamefont {C.}~\bibnamefont
    {Gross}},\ }\bibfield  {title} {\enquote {\bibinfo {title} {Exploring the
  many-body localization transition in two dimensions},}\ }\href {\doibase
  10.1126/science.aaf8834} {\bibfield  {journal} {\bibinfo  {journal}
  {Science}\ }\textbf {\bibinfo {volume} {352}},\ \bibinfo {pages} {1547--1552}
(\bibinfo {year} {2016})}\BibitemShut {NoStop}%
\bibitem [{\citenamefont {Bordia}\ \emph
  {et~al.}(2017{\natexlab{a}})\citenamefont {Bordia}, \citenamefont
  {L{\"u}schen}, \citenamefont {Scherg}, \citenamefont {Gopalakrishnan},
  \citenamefont {Knap}, \citenamefont {Schneider},\ and\ \citenamefont
{Bloch}}]{BordiaBloch2017}%
\BibitemOpen
\bibfield  {author} {\bibinfo {author} {\bibfnamefont {P.}~\bibnamefont
  {Bordia}}, \bibinfo {author} {\bibfnamefont {H.}~\bibnamefont {L{\"u}schen}},
  \bibinfo {author} {\bibfnamefont {S.}~\bibnamefont {Scherg}}, \bibinfo
  {author} {\bibfnamefont {S.}~\bibnamefont {Gopalakrishnan}}, \bibinfo
  {author} {\bibfnamefont {M.}~\bibnamefont {Knap}}, \bibinfo {author}
  {\bibfnamefont {U.}~\bibnamefont {Schneider}}, \ and\ \bibinfo {author}
  {\bibfnamefont {I.}~\bibnamefont {Bloch}},\ }\bibfield  {title} {\enquote
  {\bibinfo {title} {Probing {{Slow Relaxation}} and {{Many}}-{{Body
Localization}} in {{Two}}-{{Dimensional Quasiperiodic Systems}}},}\ }\href
{\doibase 10.1103/PhysRevX.7.041047} {\bibfield  {journal} {\bibinfo
  {journal} {Phys. Rev. X}\ }\textbf {\bibinfo {volume} {7}},\ \bibinfo {pages}
{041047} (\bibinfo {year} {2017}{\natexlab{a}})}\BibitemShut {NoStop}%
\bibitem [{\citenamefont {Bordia}\ \emph
  {et~al.}(2017{\natexlab{b}})\citenamefont {Bordia}, \citenamefont
  {L{\"u}schen}, \citenamefont {Schneider}, \citenamefont {Knap},\ and\
\citenamefont {Bloch}}]{BordiaBloch2017a}%
\BibitemOpen
\bibfield  {author} {\bibinfo {author} {\bibfnamefont {P.}~\bibnamefont
  {Bordia}}, \bibinfo {author} {\bibfnamefont {H.}~\bibnamefont {L{\"u}schen}},
  \bibinfo {author} {\bibfnamefont {U.}~\bibnamefont {Schneider}}, \bibinfo
  {author} {\bibfnamefont {M.}~\bibnamefont {Knap}}, \ and\ \bibinfo {author}
  {\bibfnamefont {I.}~\bibnamefont {Bloch}},\ }\bibfield  {title} {\enquote
  {\bibinfo {title} {Periodically driving a many-body localized quantum
  system},}\ }\href {\doibase 10.1038/nphys4020} {\bibfield  {journal}
  {\bibinfo  {journal} {Nat. Phys.}\ }\textbf {\bibinfo {volume} {13}},\
  \bibinfo {pages} {460--464} (\bibinfo {year}
{2017}{\natexlab{b}})}\BibitemShut {NoStop}%
\bibitem [{\citenamefont {L{\"u}schen}\ \emph {et~al.}(2017)\citenamefont
  {L{\"u}schen}, \citenamefont {Bordia}, \citenamefont {Hodgman}, \citenamefont
  {Schreiber}, \citenamefont {Sarkar}, \citenamefont {Daley}, \citenamefont
  {Fischer}, \citenamefont {Altman}, \citenamefont {Bloch},\ and\ \citenamefont
{Schneider}}]{LuschenSchneider2017}%
\BibitemOpen
\bibfield  {author} {\bibinfo {author} {\bibfnamefont {H.~P.}\ \bibnamefont
  {L{\"u}schen}}, \bibinfo {author} {\bibfnamefont {P.}~\bibnamefont {Bordia}},
  \bibinfo {author} {\bibfnamefont {S.~S.}\ \bibnamefont {Hodgman}}, \bibinfo
  {author} {\bibfnamefont {M.}~\bibnamefont {Schreiber}}, \bibinfo {author}
  {\bibfnamefont {S.}~\bibnamefont {Sarkar}}, \bibinfo {author} {\bibfnamefont
    {A.~J.}\ \bibnamefont {Daley}}, \bibinfo {author} {\bibfnamefont {M.~H.}\
    \bibnamefont {Fischer}}, \bibinfo {author} {\bibfnamefont {E.}~\bibnamefont
  {Altman}}, \bibinfo {author} {\bibfnamefont {I.}~\bibnamefont {Bloch}}, \
  and\ \bibinfo {author} {\bibfnamefont {U.}~\bibnamefont {Schneider}},\
  }\bibfield  {title} {\enquote {\bibinfo {title} {Signatures of
  {{Many}}-{{Body Localization}} in a {{Controlled Open Quantum System}}},}\
  }\href {\doibase 10.1103/PhysRevX.7.011034} {\bibfield  {journal} {\bibinfo
  {journal} {Phys. Rev. X}\ }\textbf {\bibinfo {volume} {7}},\ \bibinfo {pages}
{011034} (\bibinfo {year} {2017})}\BibitemShut {NoStop}%
\bibitem [{\citenamefont {Smith}\ \emph {et~al.}(2016)\citenamefont {Smith},
  \citenamefont {Lee}, \citenamefont {Richerme}, \citenamefont {Neyenhuis},
  \citenamefont {Hess}, \citenamefont {Hauke}, \citenamefont {Heyl},
\citenamefont {Huse},\ and\ \citenamefont {Monroe}}]{SmithMonroe2016}%
\BibitemOpen
\bibfield  {author} {\bibinfo {author} {\bibfnamefont {J.}~\bibnamefont
  {Smith}}, \bibinfo {author} {\bibfnamefont {A.}~\bibnamefont {Lee}}, \bibinfo
  {author} {\bibfnamefont {P.}~\bibnamefont {Richerme}}, \bibinfo {author}
  {\bibfnamefont {B.}~\bibnamefont {Neyenhuis}}, \bibinfo {author}
  {\bibfnamefont {P.~W.}\ \bibnamefont {Hess}}, \bibinfo {author}
  {\bibfnamefont {P.}~\bibnamefont {Hauke}}, \bibinfo {author} {\bibfnamefont
    {M.}~\bibnamefont {Heyl}}, \bibinfo {author} {\bibfnamefont {D.~A.}\
    \bibnamefont {Huse}}, \ and\ \bibinfo {author} {\bibfnamefont
    {C.}~\bibnamefont {Monroe}},\ }\bibfield  {title} {\enquote {\bibinfo {title}
    {Many-body localization in a quantum simulator with programmable random
  disorder},}\ }\href {\doibase 10.1038/nphys3783} {\bibfield  {journal}
  {\bibinfo  {journal} {Nat. Phys.}\ }\textbf {\bibinfo {volume} {12}},\
\bibinfo {pages} {907--911} (\bibinfo {year} {2016})}\BibitemShut {NoStop}%
\bibitem [{\citenamefont {Roushan}\ \emph {et~al.}(2017)\citenamefont
  {Roushan}, \citenamefont {Neill}, \citenamefont {Tangpanitanon},
  \citenamefont {Bastidas}, \citenamefont {Megrant}, \citenamefont {Barends},
  \citenamefont {Chen}, \citenamefont {Chen}, \citenamefont {Chiaro},
  \citenamefont {Dunsworth}, \citenamefont {Fowler}, \citenamefont {Foxen},
  \citenamefont {Giustina}, \citenamefont {Jeffrey}, \citenamefont {Kelly},
  \citenamefont {Lucero}, \citenamefont {Mutus}, \citenamefont {Neeley},
  \citenamefont {Quintana}, \citenamefont {Sank}, \citenamefont {Vainsencher},
  \citenamefont {Wenner}, \citenamefont {White}, \citenamefont {Neven},
  \citenamefont {Angelakis},\ and\ \citenamefont
{Martinis}}]{RoushanMartinis2017}%
\BibitemOpen
\bibfield  {author} {\bibinfo {author} {\bibfnamefont {P.}~\bibnamefont
  {Roushan}}, \bibinfo {author} {\bibfnamefont {C.}~\bibnamefont {Neill}},
  \bibinfo {author} {\bibfnamefont {J.}~\bibnamefont {Tangpanitanon}}, \bibinfo
  {author} {\bibfnamefont {V.~M.}\ \bibnamefont {Bastidas}}, \bibinfo {author}
  {\bibfnamefont {A.}~\bibnamefont {Megrant}}, \bibinfo {author} {\bibfnamefont
    {R.}~\bibnamefont {Barends}}, \bibinfo {author} {\bibfnamefont
    {Y.}~\bibnamefont {Chen}}, \bibinfo {author} {\bibfnamefont {Z.}~\bibnamefont
  {Chen}}, \bibinfo {author} {\bibfnamefont {B.}~\bibnamefont {Chiaro}},
  \bibinfo {author} {\bibfnamefont {A.}~\bibnamefont {Dunsworth}}, \bibinfo
  {author} {\bibfnamefont {A.}~\bibnamefont {Fowler}}, \bibinfo {author}
  {\bibfnamefont {B.}~\bibnamefont {Foxen}}, \bibinfo {author} {\bibfnamefont
    {M.}~\bibnamefont {Giustina}}, \bibinfo {author} {\bibfnamefont
    {E.}~\bibnamefont {Jeffrey}}, \bibinfo {author} {\bibfnamefont
    {J.}~\bibnamefont {Kelly}}, \bibinfo {author} {\bibfnamefont
    {E.}~\bibnamefont {Lucero}}, \bibinfo {author} {\bibfnamefont
    {J.}~\bibnamefont {Mutus}}, \bibinfo {author} {\bibfnamefont
    {M.}~\bibnamefont {Neeley}}, \bibinfo {author} {\bibfnamefont
    {C.}~\bibnamefont {Quintana}}, \bibinfo {author} {\bibfnamefont
    {D.}~\bibnamefont {Sank}}, \bibinfo {author} {\bibfnamefont {A.}~\bibnamefont
  {Vainsencher}}, \bibinfo {author} {\bibfnamefont {J.}~\bibnamefont {Wenner}},
  \bibinfo {author} {\bibfnamefont {T.}~\bibnamefont {White}}, \bibinfo
  {author} {\bibfnamefont {H.}~\bibnamefont {Neven}}, \bibinfo {author}
  {\bibfnamefont {D.~G.}\ \bibnamefont {Angelakis}}, \ and\ \bibinfo {author}
  {\bibfnamefont {J.}~\bibnamefont {Martinis}},\ }\bibfield  {title} {\enquote
  {\bibinfo {title} {Spectroscopic signatures of localization with interacting
  photons in superconducting qubits},}\ }\href {\doibase
  10.1126/science.aao1401} {\bibfield  {journal} {\bibinfo  {journal}
  {Science}\ }\textbf {\bibinfo {volume} {358}},\ \bibinfo {pages} {1175--1179}
(\bibinfo {year} {2017})}\BibitemShut {NoStop}%
\bibitem [{\citenamefont {Guo}\ \emph {et~al.}(2020)\citenamefont {Guo},
  \citenamefont {Cheng}, \citenamefont {Sun}, \citenamefont {Song},
  \citenamefont {Li}, \citenamefont {Wang}, \citenamefont {Ren}, \citenamefont
  {Dong}, \citenamefont {Zheng}, \citenamefont {Zhang}, \citenamefont
{Mondaini}, \citenamefont {Fan},\ and\ \citenamefont {Wang}}]{GuoWang2020}%
\BibitemOpen
\bibfield  {author} {\bibinfo {author} {\bibfnamefont {Q.}~\bibnamefont
  {Guo}}, \bibinfo {author} {\bibfnamefont {C.}~\bibnamefont {Cheng}}, \bibinfo
  {author} {\bibfnamefont {Z.-H.}\ \bibnamefont {Sun}}, \bibinfo {author}
  {\bibfnamefont {Z.}~\bibnamefont {Song}}, \bibinfo {author} {\bibfnamefont
    {H.}~\bibnamefont {Li}}, \bibinfo {author} {\bibfnamefont {Z.}~\bibnamefont
  {Wang}}, \bibinfo {author} {\bibfnamefont {W.}~\bibnamefont {Ren}}, \bibinfo
  {author} {\bibfnamefont {H.}~\bibnamefont {Dong}}, \bibinfo {author}
  {\bibfnamefont {D.}~\bibnamefont {Zheng}}, \bibinfo {author} {\bibfnamefont
    {Y.-R.}\ \bibnamefont {Zhang}}, \bibinfo {author} {\bibfnamefont
    {R.}~\bibnamefont {Mondaini}}, \bibinfo {author} {\bibfnamefont
    {H.}~\bibnamefont {Fan}}, \ and\ \bibinfo {author} {\bibfnamefont
    {H.}~\bibnamefont {Wang}},\ }\bibfield  {title} {\enquote {\bibinfo {title}
  {Observation of energy-resolved many-body localization},}\ }\href {\doibase
  10.1038/s41567-020-1035-1} {\bibfield  {journal} {\bibinfo  {journal} {Nat.
Phys.}\ ,\ \bibinfo {pages} {1--6}} (\bibinfo {year} {2020})}\BibitemShut
{NoStop}%
\bibitem [{\citenamefont {Nandkishore}, \citenamefont {Gopalakrishnan},\ and\
\citenamefont {Huse}(2014)}]{NandkishoreHuse2014}%
\BibitemOpen
\bibfield  {author} {\bibinfo {author} {\bibfnamefont {R.}~\bibnamefont
    {Nandkishore}}, \bibinfo {author} {\bibfnamefont {S.}~\bibnamefont
    {Gopalakrishnan}}, \ and\ \bibinfo {author} {\bibfnamefont {D.~A.}\
    \bibnamefont {Huse}},\ }\bibfield  {title} {\enquote {\bibinfo {title}
    {Spectral features of a many-body-localized system weakly coupled to a
  bath},}\ }\href {\doibase 10.1103/PhysRevB.90.064203} {\bibfield  {journal}
  {\bibinfo  {journal} {Phys. Rev. B}\ }\textbf {\bibinfo {volume} {90}},\
\bibinfo {pages} {064203} (\bibinfo {year} {2014})}\BibitemShut {NoStop}%
\bibitem [{\citenamefont {Johri}, \citenamefont {Nandkishore},\ and\
\citenamefont {Bhatt}(2015)}]{JohriBhatt2015}%
\BibitemOpen
\bibfield  {author} {\bibinfo {author} {\bibfnamefont {S.}~\bibnamefont
  {Johri}}, \bibinfo {author} {\bibfnamefont {R.}~\bibnamefont {Nandkishore}},
  \ and\ \bibinfo {author} {\bibfnamefont {R.~N.}\ \bibnamefont {Bhatt}},\
  }\bibfield  {title} {\enquote {\bibinfo {title} {Many-{{Body Localization}}
  in {{Imperfectly Isolated Quantum Systems}}},}\ }\href {\doibase
  10.1103/PhysRevLett.114.117401} {\bibfield  {journal} {\bibinfo  {journal}
  {Phys. Rev. Lett.}\ }\textbf {\bibinfo {volume} {114}},\ \bibinfo {pages}
{117401} (\bibinfo {year} {2015})}\BibitemShut {NoStop}%
\bibitem [{\citenamefont {Nandkishore}(2015)}]{Nandkishore2015}%
\BibitemOpen
\bibfield  {author} {\bibinfo {author} {\bibfnamefont {R.}~\bibnamefont
    {Nandkishore}},\ }\bibfield  {title} {\enquote {\bibinfo {title} {Many-body
  localization proximity effect},}\ }\href {\doibase
  10.1103/PhysRevB.92.245141} {\bibfield  {journal} {\bibinfo  {journal} {Phys.
  Rev. B}\ }\textbf {\bibinfo {volume} {92}},\ \bibinfo {pages} {245141}
(\bibinfo {year} {2015})}\BibitemShut {NoStop}%
\bibitem [{\citenamefont {{\v Z}nidari{\v c}}, \citenamefont {Scardicchio},\
and\ \citenamefont {Varma}(2016)}]{ZnidaricVarma2016}%
\BibitemOpen
\bibfield  {author} {\bibinfo {author} {\bibfnamefont {M.}~\bibnamefont {{\v
    Z}nidari{\v c}}}, \bibinfo {author} {\bibfnamefont {A.}~\bibnamefont
    {Scardicchio}}, \ and\ \bibinfo {author} {\bibfnamefont {V.~K.}\ \bibnamefont
    {Varma}},\ }\bibfield  {title} {\enquote {\bibinfo {title} {Diffusive and
      subdiffusive spin transport in the ergodic phase of a many-body localizable
      system},}\ }\href@noop {} {\bibfield  {journal} {\bibinfo  {journal} {Phys.
  Rev. Lett.}\ }\textbf {\bibinfo {volume} {117}},\ \bibinfo {pages} {040601}
(\bibinfo {year} {2016})}\BibitemShut {NoStop}%
\bibitem [{\citenamefont {Medvedyeva}, \citenamefont {Prosen},\ and\
\citenamefont {{\v Z}nidari{\v c}}(2016)}]{MedvedyevaZnidaric2016}%
\BibitemOpen
\bibfield  {author} {\bibinfo {author} {\bibfnamefont {M.~V.}\ \bibnamefont
  {Medvedyeva}}, \bibinfo {author} {\bibfnamefont {T.}~\bibnamefont {Prosen}},
  \ and\ \bibinfo {author} {\bibfnamefont {M.}~\bibnamefont {{\v Z}nidari{\v
      c}}},\ }\bibfield  {title} {\enquote {\bibinfo {title} {Influence of
  dephasing on many-body localization},}\ }\href {\doibase
  10.1103/PhysRevB.93.094205} {\bibfield  {journal} {\bibinfo  {journal} {Phys.
  Rev. B}\ }\textbf {\bibinfo {volume} {93}},\ \bibinfo {pages} {094205}
(\bibinfo {year} {2016})}\BibitemShut {NoStop}%
\bibitem [{\citenamefont {Levi}\ \emph {et~al.}(2016)\citenamefont {Levi},
  \citenamefont {Heyl}, \citenamefont {Lesanovsky},\ and\ \citenamefont
{Garrahan}}]{LeviGarrahan2016}%
\BibitemOpen
\bibfield  {author} {\bibinfo {author} {\bibfnamefont {E.}~\bibnamefont
  {Levi}}, \bibinfo {author} {\bibfnamefont {M.}~\bibnamefont {Heyl}}, \bibinfo
  {author} {\bibfnamefont {I.}~\bibnamefont {Lesanovsky}}, \ and\ \bibinfo
{author} {\bibfnamefont {J.~P.}\ \bibnamefont {Garrahan}},\ }\bibfield
{title} {\enquote {\bibinfo {title} {Robustness of {{Many}}-{{Body
  Localization}} in the {{Presence}} of {{Dissipation}}},}\ }\href {\doibase
  10.1103/PhysRevLett.116.237203} {\bibfield  {journal} {\bibinfo  {journal}
  {Phys. Rev. Lett.}\ }\textbf {\bibinfo {volume} {116}},\ \bibinfo {pages}
{237203} (\bibinfo {year} {2016})}\BibitemShut {NoStop}%
\bibitem [{\citenamefont {Fischer}, \citenamefont {Maksymenko},\ and\
\citenamefont {Altman}(2016)}]{FischerAltman2016}%
\BibitemOpen
\bibfield  {author} {\bibinfo {author} {\bibfnamefont {M.~H.}\ \bibnamefont
  {Fischer}}, \bibinfo {author} {\bibfnamefont {M.}~\bibnamefont {Maksymenko}},
  \ and\ \bibinfo {author} {\bibfnamefont {E.}~\bibnamefont {Altman}},\
  }\bibfield  {title} {\enquote {\bibinfo {title} {Dynamics of a
{{Many}}-{{Body}}-{{Localized System Coupled}} to a {{Bath}}},}\ }\href
{\doibase 10.1103/PhysRevLett.116.160401} {\bibfield  {journal} {\bibinfo
  {journal} {Phys. Rev. Lett.}\ }\textbf {\bibinfo {volume} {116}},\ \bibinfo
{pages} {160401} (\bibinfo {year} {2016})}\BibitemShut {NoStop}%
\bibitem [{\citenamefont {Hyatt}\ \emph {et~al.}(2017)\citenamefont {Hyatt},
  \citenamefont {Garrison}, \citenamefont {Potter},\ and\ \citenamefont
{Bauer}}]{HyattBauer2017}%
\BibitemOpen
\bibfield  {author} {\bibinfo {author} {\bibfnamefont {K.}~\bibnamefont
  {Hyatt}}, \bibinfo {author} {\bibfnamefont {J.~R.}\ \bibnamefont {Garrison}},
  \bibinfo {author} {\bibfnamefont {A.~C.}\ \bibnamefont {Potter}}, \ and\
\bibinfo {author} {\bibfnamefont {B.}~\bibnamefont {Bauer}},\ }\bibfield
{title} {\enquote {\bibinfo {title} {Many-body localization in the presence
  of a small bath},}\ }\href {\doibase 10.1103/PhysRevB.95.035132} {\bibfield
  {journal} {\bibinfo  {journal} {Phys. Rev. B}\ }\textbf {\bibinfo {volume}
{95}},\ \bibinfo {pages} {035132} (\bibinfo {year} {2017})}\BibitemShut
{NoStop}%
\bibitem [{\citenamefont {Everest}\ \emph {et~al.}(2017)\citenamefont
  {Everest}, \citenamefont {Lesanovsky}, \citenamefont {Garrahan},\ and\
\citenamefont {Levi}}]{EverestLevi2017}%
\BibitemOpen
\bibfield  {author} {\bibinfo {author} {\bibfnamefont {B.}~\bibnamefont
  {Everest}}, \bibinfo {author} {\bibfnamefont {I.}~\bibnamefont {Lesanovsky}},
  \bibinfo {author} {\bibfnamefont {J.~P.}\ \bibnamefont {Garrahan}}, \ and\
\bibinfo {author} {\bibfnamefont {E.}~\bibnamefont {Levi}},\ }\bibfield
{title} {\enquote {\bibinfo {title} {Role of interactions in a dissipative
many-body localized system},}\ }\href {\doibase 10.1103/PhysRevB.95.024310}
{\bibfield  {journal} {\bibinfo  {journal} {Phys. Rev. B}\ }\textbf {\bibinfo
  {volume} {95}},\ \bibinfo {pages} {024310} (\bibinfo {year}
{2017})}\BibitemShut {NoStop}%
\bibitem [{\citenamefont {Luitz}, \citenamefont {Huveneers},\ and\
\citenamefont {De~Roeck}(2017)}]{LuitzDeRoeck2017}%
\BibitemOpen
\bibfield  {author} {\bibinfo {author} {\bibfnamefont {D.~J.}\ \bibnamefont
  {Luitz}}, \bibinfo {author} {\bibfnamefont {F.}~\bibnamefont {Huveneers}}, \
  and\ \bibinfo {author} {\bibfnamefont {W.}~\bibnamefont {De~Roeck}},\
  }\bibfield  {title} {\enquote {\bibinfo {title} {How a {{Small Quantum Bath
  Can Thermalize Long Localized Chains}}},}\ }\href {\doibase
  10.1103/PhysRevLett.119.150602} {\bibfield  {journal} {\bibinfo  {journal}
  {Phys. Rev. Lett.}\ }\textbf {\bibinfo {volume} {119}},\ \bibinfo {pages}
{150602} (\bibinfo {year} {2017})}\BibitemShut {NoStop}%
\bibitem [{\citenamefont {Nandkishore}\ and\ \citenamefont
{Gopalakrishnan}(2017)}]{NandkishoreGopalakrishnan2017}%
\BibitemOpen
\bibfield  {author} {\bibinfo {author} {\bibfnamefont {R.}~\bibnamefont
    {Nandkishore}}\ and\ \bibinfo {author} {\bibfnamefont {S.}~\bibnamefont
    {Gopalakrishnan}},\ }\bibfield  {title} {\enquote {\bibinfo {title} {Many
      body localized systems weakly coupled to baths: {{Many}} body localized
systems weakly coupled to baths},}\ }\href {\doibase 10.1002/andp.201600181}
{\bibfield  {journal} {\bibinfo  {journal} {Ann. Phys. (Berlin)}\ }\textbf
  {\bibinfo {volume} {529}},\ \bibinfo {pages} {1600181} (\bibinfo {year}
{2017})}\BibitemShut {NoStop}%
\bibitem [{\citenamefont {{\v Z}nidari{\v c}}\ \emph
  {et~al.}(2017)\citenamefont {{\v Z}nidari{\v c}}, \citenamefont
  {Mendoza-Arenas}, \citenamefont {Clark},\ and\ \citenamefont
{Goold}}]{ZnidaricGoold2017}%
\BibitemOpen
\bibfield  {author} {\bibinfo {author} {\bibfnamefont {M.}~\bibnamefont {{\v
    Z}nidari{\v c}}}, \bibinfo {author} {\bibfnamefont {J.~J.}\ \bibnamefont
    {Mendoza-Arenas}}, \bibinfo {author} {\bibfnamefont {S.~R.}\ \bibnamefont
    {Clark}}, \ and\ \bibinfo {author} {\bibfnamefont {J.}~\bibnamefont
    {Goold}},\ }\bibfield  {title} {\enquote {\bibinfo {title} {Dephasing
      enhanced spin transport in the ergodic phase of a many-body localizable
  system},}\ }\href {\doibase 10.1002/andp.201600298} {\bibfield  {journal}
  {\bibinfo  {journal} {Ann. Phys. (Berlin)}\ }\textbf {\bibinfo {volume}
{529}},\ \bibinfo {pages} {1600298} (\bibinfo {year} {2017})}\BibitemShut
{NoStop}%
\bibitem [{\citenamefont {{\v Z}nidari{\v c}}\ and\ \citenamefont
{Ljubotina}(2018)}]{ZnidaricLjubotina2018}%
\BibitemOpen
\bibfield  {author} {\bibinfo {author} {\bibfnamefont {M.}~\bibnamefont {{\v
    Z}nidari{\v c}}}\ and\ \bibinfo {author} {\bibfnamefont {M.}~\bibnamefont
    {Ljubotina}},\ }\bibfield  {title} {\enquote {\bibinfo {title} {Interaction
  instability of localization in quasiperiodic systems},}\ }\href {\doibase
  10.1073/pnas.1800589115} {\bibfield  {journal} {\bibinfo  {journal} {PNAS}\
  }\textbf {\bibinfo {volume} {115}},\ \bibinfo {pages} {4595--4600} (\bibinfo
{year} {2018})}\BibitemShut {NoStop}%
\bibitem [{\citenamefont {Karlsson}, \citenamefont {Hopjan},\ and\
\citenamefont {Verdozzi}(2018)}]{KarlssonVerdozzi2018}%
\BibitemOpen
\bibfield  {author} {\bibinfo {author} {\bibfnamefont {D.}~\bibnamefont
  {Karlsson}}, \bibinfo {author} {\bibfnamefont {M.}~\bibnamefont {Hopjan}}, \
  and\ \bibinfo {author} {\bibfnamefont {C.}~\bibnamefont {Verdozzi}},\
  }\bibfield  {title} {\enquote {\bibinfo {title} {Disorder and interactions in
      systems out of equilibrium: {{The}} exact independent-particle picture from
density functional theory},}\ }\href {\doibase 10.1103/PhysRevB.97.125151}
{\bibfield  {journal} {\bibinfo  {journal} {Phys. Rev. B}\ }\textbf {\bibinfo
  {volume} {97}},\ \bibinfo {pages} {125151} (\bibinfo {year}
{2018})}\BibitemShut {NoStop}%
\bibitem [{\citenamefont {Vakulchyk}\ \emph {et~al.}(2018)\citenamefont
  {Vakulchyk}, \citenamefont {Yusipov}, \citenamefont {Ivanchenko},
\citenamefont {Flach},\ and\ \citenamefont {Denisov}}]{VakulchykDenisov2018}%
\BibitemOpen
\bibfield  {author} {\bibinfo {author} {\bibfnamefont {I.}~\bibnamefont
  {Vakulchyk}}, \bibinfo {author} {\bibfnamefont {I.}~\bibnamefont {Yusipov}},
  \bibinfo {author} {\bibfnamefont {M.}~\bibnamefont {Ivanchenko}}, \bibinfo
  {author} {\bibfnamefont {S.}~\bibnamefont {Flach}}, \ and\ \bibinfo {author}
  {\bibfnamefont {S.}~\bibnamefont {Denisov}},\ }\bibfield  {title} {\enquote
  {\bibinfo {title} {Signatures of many-body localization in steady states of
open quantum systems},}\ }\href {\doibase 10.1103/PhysRevB.98.020202}
{\bibfield  {journal} {\bibinfo  {journal} {Phys. Rev. B}\ }\textbf {\bibinfo
  {volume} {98}},\ \bibinfo {pages} {020202} (\bibinfo {year}
{2018})}\BibitemShut {NoStop}%
\bibitem [{\citenamefont {Xu}, \citenamefont {Guo},\ and\ \citenamefont
{Poletti}(2018)}]{XuPoletti2018}%
\BibitemOpen
\bibfield  {author} {\bibinfo {author} {\bibfnamefont {X.}~\bibnamefont
  {Xu}}, \bibinfo {author} {\bibfnamefont {C.}~\bibnamefont {Guo}}, \ and\
\bibinfo {author} {\bibfnamefont {D.}~\bibnamefont {Poletti}},\ }\bibfield
{title} {\enquote {\bibinfo {title} {Interplay of interaction and disorder in
  the steady state of an open quantum system},}\ }\href {\doibase
  10.1103/PhysRevB.97.140201} {\bibfield  {journal} {\bibinfo  {journal} {Phys.
  Rev. B}\ }\textbf {\bibinfo {volume} {97}},\ \bibinfo {pages} {140201}
(\bibinfo {year} {2018})}\BibitemShut {NoStop}%
\bibitem [{\citenamefont {Wu}\ \emph {et~al.}(2019)\citenamefont {Wu},
  \citenamefont {Schnell}, \citenamefont {Tomasi}, \citenamefont {Heyl},\ and\
\citenamefont {Eckardt}}]{WuEckardt2019a}%
\BibitemOpen
\bibfield  {author} {\bibinfo {author} {\bibfnamefont {L.-N.}\ \bibnamefont
  {Wu}}, \bibinfo {author} {\bibfnamefont {A.}~\bibnamefont {Schnell}},
  \bibinfo {author} {\bibfnamefont {G.~D.}\ \bibnamefont {Tomasi}}, \bibinfo
  {author} {\bibfnamefont {M.}~\bibnamefont {Heyl}}, \ and\ \bibinfo {author}
  {\bibfnamefont {A.}~\bibnamefont {Eckardt}},\ }\bibfield  {title} {\enquote
  {\bibinfo {title} {Describing many-body localized systems in thermal
  environments},}\ }\href {\doibase 10.1088/1367-2630/ab25a4} {\bibfield
  {journal} {\bibinfo  {journal} {New J. Phys.}\ }\textbf {\bibinfo {volume}
{21}},\ \bibinfo {pages} {063026} (\bibinfo {year} {2019})}\BibitemShut
{NoStop}%
\bibitem [{\citenamefont {{Rubio-Abadal}}\ \emph {et~al.}(2019)\citenamefont
  {{Rubio-Abadal}}, \citenamefont {Choi}, \citenamefont {Zeiher}, \citenamefont
  {Hollerith}, \citenamefont {Rui}, \citenamefont {Bloch},\ and\ \citenamefont
{Gross}}]{Rubio-AbadalGross2019}%
\BibitemOpen
\bibfield  {author} {\bibinfo {author} {\bibfnamefont {A.}~\bibnamefont
    {{Rubio-Abadal}}}, \bibinfo {author} {\bibfnamefont {J.-y.}\ \bibnamefont
  {Choi}}, \bibinfo {author} {\bibfnamefont {J.}~\bibnamefont {Zeiher}},
  \bibinfo {author} {\bibfnamefont {S.}~\bibnamefont {Hollerith}}, \bibinfo
  {author} {\bibfnamefont {J.}~\bibnamefont {Rui}}, \bibinfo {author}
  {\bibfnamefont {I.}~\bibnamefont {Bloch}}, \ and\ \bibinfo {author}
  {\bibfnamefont {C.}~\bibnamefont {Gross}},\ }\bibfield  {title} {\enquote
  {\bibinfo {title} {Many-{{Body Delocalization}} in the {{Presence}} of a
  {{Quantum Bath}}},}\ }\href {\doibase 10.1103/PhysRevX.9.041014} {\bibfield
  {journal} {\bibinfo  {journal} {Phys. Rev. X}\ }\textbf {\bibinfo {volume}
{9}},\ \bibinfo {pages} {041014} (\bibinfo {year} {2019})}\BibitemShut
{NoStop}%
\bibitem [{\citenamefont {Wybo}, \citenamefont {Knap},\ and\ \citenamefont
{Pollmann}(2020)}]{WyboPollmann2020a}%
\BibitemOpen
\bibfield  {author} {\bibinfo {author} {\bibfnamefont {E.}~\bibnamefont
  {Wybo}}, \bibinfo {author} {\bibfnamefont {M.}~\bibnamefont {Knap}}, \ and\
\bibinfo {author} {\bibfnamefont {F.}~\bibnamefont {Pollmann}},\ }\bibfield
{title} {\enquote {\bibinfo {title} {Entanglement dynamics of a many-body
  localized system coupled to a bath},}\ }\href {\doibase
  10.1103/PhysRevB.102.064304} {\bibfield  {journal} {\bibinfo  {journal}
  {Phys. Rev. B}\ }\textbf {\bibinfo {volume} {102}},\ \bibinfo {pages}
{064304} (\bibinfo {year} {2020})}\BibitemShut {NoStop}%
\bibitem [{\citenamefont {Diehl}\ \emph {et~al.}(2008)\citenamefont {Diehl},
  \citenamefont {Micheli}, \citenamefont {Kantian}, \citenamefont {Kraus},
\citenamefont {B{\"u}chler},\ and\ \citenamefont {Zoller}}]{DiehlZoller2008}%
\BibitemOpen
\bibfield  {author} {\bibinfo {author} {\bibfnamefont {S.}~\bibnamefont
  {Diehl}}, \bibinfo {author} {\bibfnamefont {A.}~\bibnamefont {Micheli}},
  \bibinfo {author} {\bibfnamefont {A.}~\bibnamefont {Kantian}}, \bibinfo
  {author} {\bibfnamefont {B.}~\bibnamefont {Kraus}}, \bibinfo {author}
  {\bibfnamefont {H.~P.}\ \bibnamefont {B{\"u}chler}}, \ and\ \bibinfo {author}
  {\bibfnamefont {P.}~\bibnamefont {Zoller}},\ }\bibfield  {title} {\enquote
  {\bibinfo {title} {Quantum states and phases in driven open quantum systems
  with cold atoms},}\ }\href {\doibase 10.1038/nphys1073} {\bibfield  {journal}
  {\bibinfo  {journal} {Nat. Phys.}\ }\textbf {\bibinfo {volume} {4}},\
\bibinfo {pages} {878--883} (\bibinfo {year} {2008})}\BibitemShut {NoStop}%
\bibitem [{\citenamefont {Schindler}\ \emph {et~al.}(2013)\citenamefont
  {Schindler}, \citenamefont {M{\"u}ller}, \citenamefont {Nigg}, \citenamefont
  {Barreiro}, \citenamefont {Martinez}, \citenamefont {Hennrich}, \citenamefont
  {Monz}, \citenamefont {Diehl}, \citenamefont {Zoller},\ and\ \citenamefont
{Blatt}}]{SchindlerBlatt2013}%
\BibitemOpen
\bibfield  {author} {\bibinfo {author} {\bibfnamefont {P.}~\bibnamefont
    {Schindler}}, \bibinfo {author} {\bibfnamefont {M.}~\bibnamefont
  {M{\"u}ller}}, \bibinfo {author} {\bibfnamefont {D.}~\bibnamefont {Nigg}},
  \bibinfo {author} {\bibfnamefont {J.~T.}\ \bibnamefont {Barreiro}}, \bibinfo
  {author} {\bibfnamefont {E.~A.}\ \bibnamefont {Martinez}}, \bibinfo {author}
  {\bibfnamefont {M.}~\bibnamefont {Hennrich}}, \bibinfo {author}
  {\bibfnamefont {T.}~\bibnamefont {Monz}}, \bibinfo {author} {\bibfnamefont
    {S.}~\bibnamefont {Diehl}}, \bibinfo {author} {\bibfnamefont
    {P.}~\bibnamefont {Zoller}}, \ and\ \bibinfo {author} {\bibfnamefont
    {R.}~\bibnamefont {Blatt}},\ }\bibfield  {title} {\enquote {\bibinfo {title}
  {Quantum simulation of dynamical maps with trapped ions},}\ }\href {\doibase
  10.1038/nphys2630} {\bibfield  {journal} {\bibinfo  {journal} {Nat. Phys.}\
  }\textbf {\bibinfo {volume} {9}},\ \bibinfo {pages} {361--367} (\bibinfo
{year} {2013})}\BibitemShut {NoStop}%
\bibitem [{\citenamefont {Jordan}\ and\ \citenamefont
{Wigner}(1928)}]{JordanWigner1928}%
\BibitemOpen
\bibfield  {author} {\bibinfo {author} {\bibfnamefont {P.}~\bibnamefont
    {Jordan}}\ and\ \bibinfo {author} {\bibfnamefont {E.}~\bibnamefont
    {Wigner}},\ }\bibfield  {title} {\enquote {\bibinfo {title} {{\"Uber das
Paulische \"Aquivalenzverbot}},}\ }\href {\doibase 10.1007/BF01331938}
{\bibfield  {journal} {\bibinfo  {journal} {Z. Physik}\ }\textbf {\bibinfo
  {volume} {47}},\ \bibinfo {pages} {631--651} (\bibinfo {year}
{1928})}\BibitemShut {NoStop}%
\bibitem [{\citenamefont {Lieb}, \citenamefont {Schultz},\ and\ \citenamefont
{Mattis}(1961)}]{LiebMattis1961}%
\BibitemOpen
\bibfield  {author} {\bibinfo {author} {\bibfnamefont {E.}~\bibnamefont
  {Lieb}}, \bibinfo {author} {\bibfnamefont {T.}~\bibnamefont {Schultz}}, \
  and\ \bibinfo {author} {\bibfnamefont {D.}~\bibnamefont {Mattis}},\
  }\bibfield  {title} {\enquote {\bibinfo {title} {Two soluble models of an
antiferromagnetic chain},}\ }\href {\doibase 10.1016/0003-4916(61)90115-4}
{\bibfield  {journal} {\bibinfo  {journal} {Ann. Phys. (N.Y.)}\ }\textbf
  {\bibinfo {volume} {16}},\ \bibinfo {pages} {407--466} (\bibinfo {year}
{1961})}\BibitemShut {NoStop}%
\bibitem [{\citenamefont {Bera}\ \emph {et~al.}(2015)\citenamefont {Bera},
  \citenamefont {Schomerus}, \citenamefont {{Heidrich-Meisner}},\ and\
\citenamefont {Bardarson}}]{BeraBardarson2015}%
\BibitemOpen
\bibfield  {author} {\bibinfo {author} {\bibfnamefont {S.}~\bibnamefont
  {Bera}}, \bibinfo {author} {\bibfnamefont {H.}~\bibnamefont {Schomerus}},
  \bibinfo {author} {\bibfnamefont {F.}~\bibnamefont {{Heidrich-Meisner}}}, \
  and\ \bibinfo {author} {\bibfnamefont {J.~H.}\ \bibnamefont {Bardarson}},\
  }\bibfield  {title} {\enquote {\bibinfo {title} {Many-{{Body Localization
  Characterized}} from a {{One}}-{{Particle Perspective}}},}\ }\href {\doibase
  10.1103/PhysRevLett.115.046603} {\bibfield  {journal} {\bibinfo  {journal}
  {Phys. Rev. Lett.}\ }\textbf {\bibinfo {volume} {115}},\ \bibinfo {pages}
{046603} (\bibinfo {year} {2015})}\BibitemShut {NoStop}%
\bibitem [{\citenamefont {Bera}\ \emph {et~al.}(2017)\citenamefont {Bera},
  \citenamefont {Martynec}, \citenamefont {Schomerus}, \citenamefont
{Heidrich-Meisner},\ and\ \citenamefont {Bardarson}}]{BeraBardarson2017}%
\BibitemOpen
\bibfield  {author} {\bibinfo {author} {\bibfnamefont {S.}~\bibnamefont
  {Bera}}, \bibinfo {author} {\bibfnamefont {T.}~\bibnamefont {Martynec}},
  \bibinfo {author} {\bibfnamefont {H.}~\bibnamefont {Schomerus}}, \bibinfo
  {author} {\bibfnamefont {F.}~\bibnamefont {Heidrich-Meisner}}, \ and\
  \bibinfo {author} {\bibfnamefont {J.~H.}\ \bibnamefont {Bardarson}},\
  }\bibfield  {title} {\enquote {\bibinfo {title} {One-particle density matrix
  characterization of many-body localization},}\ }\href {\doibase
  10.1002/andp.201600356} {\bibfield  {journal} {\bibinfo  {journal} {Ann.
  Phys. (Berlin)}\ }\textbf {\bibinfo {volume} {529}},\ \bibinfo {pages}
{1600356} (\bibinfo {year} {2017})}\BibitemShut {NoStop}%
\bibitem [{\citenamefont {Lezama}\ \emph {et~al.}(2017)\citenamefont {Lezama},
  \citenamefont {Bera}, \citenamefont {Schomerus}, \citenamefont
{{Heidrich-Meisner}},\ and\ \citenamefont {Bardarson}}]{LezamaBardarson2017}%
\BibitemOpen
\bibfield  {author} {\bibinfo {author} {\bibfnamefont {T.~L.~M.}\
    \bibnamefont {Lezama}}, \bibinfo {author} {\bibfnamefont {S.}~\bibnamefont
  {Bera}}, \bibinfo {author} {\bibfnamefont {H.}~\bibnamefont {Schomerus}},
  \bibinfo {author} {\bibfnamefont {F.}~\bibnamefont {{Heidrich-Meisner}}}, \
  and\ \bibinfo {author} {\bibfnamefont {J.~H.}\ \bibnamefont {Bardarson}},\
  }\bibfield  {title} {\enquote {\bibinfo {title} {One-particle density matrix
  occupation spectrum of many-body localized states after a global quench},}\
  }\href {\doibase 10.1103/PhysRevB.96.060202} {\bibfield  {journal} {\bibinfo
  {journal} {Phys. Rev. B}\ }\textbf {\bibinfo {volume} {96}},\ \bibinfo
{pages} {060202} (\bibinfo {year} {2017})}\BibitemShut {NoStop}%
\bibitem [{\citenamefont {Lin}\ \emph {et~al.}(2018)\citenamefont {Lin},
  \citenamefont {Sbierski}, \citenamefont {Dorfner}, \citenamefont {Karrasch},\
and\ \citenamefont {{Heidrich-Meisner}}}]{LinHeidrich-Meisner2018}%
\BibitemOpen
\bibfield  {author} {\bibinfo {author} {\bibfnamefont {S.-H.}\ \bibnamefont
  {Lin}}, \bibinfo {author} {\bibfnamefont {B.}~\bibnamefont {Sbierski}},
  \bibinfo {author} {\bibfnamefont {F.}~\bibnamefont {Dorfner}}, \bibinfo
  {author} {\bibfnamefont {C.}~\bibnamefont {Karrasch}}, \ and\ \bibinfo
{author} {\bibfnamefont {F.}~\bibnamefont {{Heidrich-Meisner}}},\ }\bibfield
{title} {\enquote {\bibinfo {title} {Many-body localization of spinless
  fermions with attractive interactions in one dimension},}\ }\href {\doibase
  10.21468/SciPostPhys.4.1.002} {\bibfield  {journal} {\bibinfo  {journal}
  {SciPost Physics}\ }\textbf {\bibinfo {volume} {4}},\ \bibinfo {pages} {002}
(\bibinfo {year} {2018})}\BibitemShut {NoStop}%
\bibitem [{\citenamefont {Luitz}, \citenamefont {Laflorencie},\ and\
\citenamefont {Alet}(2015)}]{LuitzAlet2015}%
\BibitemOpen
\bibfield  {author} {\bibinfo {author} {\bibfnamefont {D.~J.}\ \bibnamefont
  {Luitz}}, \bibinfo {author} {\bibfnamefont {N.}~\bibnamefont {Laflorencie}},
  \ and\ \bibinfo {author} {\bibfnamefont {F.}~\bibnamefont {Alet}},\
  }\bibfield  {title} {\enquote {\bibinfo {title} {Many-body localization edge
  in the random-field {{Heisenberg}} chain},}\ }\href@noop {} {\bibfield
  {journal} {\bibinfo  {journal} {Phys. Rev. B}\ }\textbf {\bibinfo {volume}
{91}},\ \bibinfo {pages} {081103} (\bibinfo {year} {2015})}\BibitemShut
{NoStop}%
\bibitem [{\citenamefont {Serbyn}, \citenamefont {Papi{\'c}},\ and\
\citenamefont {Abanin}(2015)}]{SerbynAbanin2015}%
\BibitemOpen
\bibfield  {author} {\bibinfo {author} {\bibfnamefont {M.}~\bibnamefont
  {Serbyn}}, \bibinfo {author} {\bibfnamefont {Z.}~\bibnamefont {Papi{\'c}}}, \
  and\ \bibinfo {author} {\bibfnamefont {D.~A.}\ \bibnamefont {Abanin}},\
  }\bibfield  {title} {\enquote {\bibinfo {title} {Criterion for
{{Many}}-{{Body Localization}}-{{Delocalization Phase Transition}}},}\ }\href
{\doibase 10.1103/PhysRevX.5.041047} {\bibfield  {journal} {\bibinfo
  {journal} {Phys. Rev. X}\ }\textbf {\bibinfo {volume} {5}},\ \bibinfo {pages}
{041047} (\bibinfo {year} {2015})}\BibitemShut {NoStop}%
\bibitem [{\citenamefont {{\v S}untajs}\ \emph {et~al.}(2020)\citenamefont {{\v
  S}untajs}, \citenamefont {Bon{\v c}a}, \citenamefont {Prosen},\ and\
\citenamefont {Vidmar}}]{SuntajsVidmar2020a}%
\BibitemOpen
\bibfield  {author} {\bibinfo {author} {\bibfnamefont {J.}~\bibnamefont {{\v
  S}untajs}}, \bibinfo {author} {\bibfnamefont {J.}~\bibnamefont {Bon{\v c}a}},
  \bibinfo {author} {\bibfnamefont {T.}~\bibnamefont {Prosen}}, \ and\ \bibinfo
{author} {\bibfnamefont {L.}~\bibnamefont {Vidmar}},\ }\bibfield  {title}
{\enquote {\bibinfo {title} {Quantum chaos challenges many-body
  localization},}\ }\href {\doibase 10.1103/PhysRevE.102.062144} {\bibfield
  {journal} {\bibinfo  {journal} {Phys. Rev. E}\ }\textbf {\bibinfo {volume}
{102}},\ \bibinfo {pages} {062144} (\bibinfo {year} {2020})}\BibitemShut
{NoStop}%
\bibitem [{\citenamefont {Abanin}\ \emph {et~al.}(2021)\citenamefont {Abanin},
  \citenamefont {Bardarson}, \citenamefont {De~Tomasi}, \citenamefont
  {Gopalakrishnan}, \citenamefont {Khemani}, \citenamefont {Parameswaran},
  \citenamefont {Pollmann}, \citenamefont {Potter}, \citenamefont {Serbyn},\
and\ \citenamefont {Vasseur}}]{AbaninVasseur2021}%
\BibitemOpen
\bibfield  {author} {\bibinfo {author} {\bibfnamefont {D.~A.}\ \bibnamefont
    {Abanin}}, \bibinfo {author} {\bibfnamefont {J.~H.}\ \bibnamefont
    {Bardarson}}, \bibinfo {author} {\bibfnamefont {G.}~\bibnamefont
    {De~Tomasi}}, \bibinfo {author} {\bibfnamefont {S.}~\bibnamefont
    {Gopalakrishnan}}, \bibinfo {author} {\bibfnamefont {V.}~\bibnamefont
    {Khemani}}, \bibinfo {author} {\bibfnamefont {S.~A.}\ \bibnamefont
    {Parameswaran}}, \bibinfo {author} {\bibfnamefont {F.}~\bibnamefont
    {Pollmann}}, \bibinfo {author} {\bibfnamefont {A.~C.}\ \bibnamefont
  {Potter}}, \bibinfo {author} {\bibfnamefont {M.}~\bibnamefont {Serbyn}}, \
  and\ \bibinfo {author} {\bibfnamefont {R.}~\bibnamefont {Vasseur}},\
  }\bibfield  {title} {\enquote {\bibinfo {title} {Distinguishing localization
  from chaos: {{Challenges}} in finite-size systems},}\ }\href {\doibase
  10.1016/j.aop.2021.168415} {\bibfield  {journal} {\bibinfo  {journal} {Annals
  of Physics}\ ,\ \bibinfo {pages} {168415}} (\bibinfo {year}
{2021})}\BibitemShut {NoStop}%
\bibitem [{\citenamefont {Panda}\ \emph {et~al.}(2019)\citenamefont {Panda},
  \citenamefont {Scardicchio}, \citenamefont {Schulz}, \citenamefont {Taylor},\
and\ \citenamefont {{\v Z}nidari{\v c}}}]{PandaZnidaric2019a}%
\BibitemOpen
\bibfield  {author} {\bibinfo {author} {\bibfnamefont {R.~K.}\ \bibnamefont
  {Panda}}, \bibinfo {author} {\bibfnamefont {A.}~\bibnamefont {Scardicchio}},
  \bibinfo {author} {\bibfnamefont {M.}~\bibnamefont {Schulz}}, \bibinfo
  {author} {\bibfnamefont {S.~R.}\ \bibnamefont {Taylor}}, \ and\ \bibinfo
{author} {\bibfnamefont {M.}~\bibnamefont {{\v Z}nidari{\v c}}},\ }\bibfield
{title} {\enquote {\bibinfo {title} {Can we study the many-body localisation
  transition?}}\ }\href {\doibase 10.1209/0295-5075/128/67003} {\bibfield
  {journal} {\bibinfo  {journal} {Europhys. Lett.}\ }\textbf {\bibinfo {volume}
{128}},\ \bibinfo {pages} {67003} (\bibinfo {year} {2019})}\BibitemShut
{NoStop}%
\bibitem [{\citenamefont {{Kiefer-Emmanouilidis}}\ \emph
  {et~al.}(2021)\citenamefont {{Kiefer-Emmanouilidis}}, \citenamefont
  {Unanyan}, \citenamefont {Fleischhauer},\ and\ \citenamefont
{Sirker}}]{Kiefer-EmmanouilidisSirker2021}%
\BibitemOpen
\bibfield  {author} {\bibinfo {author} {\bibfnamefont {M.}~\bibnamefont
    {{Kiefer-Emmanouilidis}}}, \bibinfo {author} {\bibfnamefont {R.}~\bibnamefont
    {Unanyan}}, \bibinfo {author} {\bibfnamefont {M.}~\bibnamefont
    {Fleischhauer}}, \ and\ \bibinfo {author} {\bibfnamefont {J.}~\bibnamefont
    {Sirker}},\ }\bibfield  {title} {\enquote {\bibinfo {title} {Slow
delocalization of particles in many-body localized phases},}\ }\href
{\doibase 10.1103/PhysRevB.103.024203} {\bibfield  {journal} {\bibinfo
  {journal} {Phys. Rev. B}\ }\textbf {\bibinfo {volume} {103}},\ \bibinfo
{pages} {024203} (\bibinfo {year} {2021})}\BibitemShut {NoStop}%
\bibitem [{\citenamefont {Gorini}, \citenamefont {Kossakowski},\ and\
\citenamefont {Sudarshan}(1976)}]{GoriniSudarshan1976}%
\BibitemOpen
\bibfield  {author} {\bibinfo {author} {\bibfnamefont {V.}~\bibnamefont
  {Gorini}}, \bibinfo {author} {\bibfnamefont {A.}~\bibnamefont {Kossakowski}},
  \ and\ \bibinfo {author} {\bibfnamefont {E.~C.~G.}\ \bibnamefont
    {Sudarshan}},\ }\bibfield  {title} {\enquote {\bibinfo {title} {Completely
  positive dynamical semigroups of {{N}}-level systems},}\ }\href {\doibase
  10.1063/1.522979} {\bibfield  {journal} {\bibinfo  {journal} {J. Math.
  Phys.}\ }\textbf {\bibinfo {volume} {17}},\ \bibinfo {pages} {821--825}
(\bibinfo {year} {1976})}\BibitemShut {NoStop}%
\bibitem [{\citenamefont {Lindblad}(1976)}]{Lindblad1976}%
\BibitemOpen
\bibfield  {author} {\bibinfo {author} {\bibfnamefont {G.}~\bibnamefont
    {Lindblad}},\ }\bibfield  {title} {\enquote {\bibinfo {title} {On the
  generators of quantum dynamical semigroups},}\ }\href {\doibase
  10.1007/BF01608499} {\bibfield  {journal} {\bibinfo  {journal} {Commun. Math.
  Phys.}\ }\textbf {\bibinfo {volume} {48}},\ \bibinfo {pages} {119--130}
(\bibinfo {year} {1976})}\BibitemShut {NoStop}%
\bibitem [{\citenamefont {Diehl}\ \emph {et~al.}(2010)\citenamefont {Diehl},
  \citenamefont {Tomadin}, \citenamefont {Micheli}, \citenamefont {Fazio},\
and\ \citenamefont {Zoller}}]{DiehlZoller2010}%
\BibitemOpen
\bibfield  {author} {\bibinfo {author} {\bibfnamefont {S.}~\bibnamefont
  {Diehl}}, \bibinfo {author} {\bibfnamefont {A.}~\bibnamefont {Tomadin}},
  \bibinfo {author} {\bibfnamefont {A.}~\bibnamefont {Micheli}}, \bibinfo
  {author} {\bibfnamefont {R.}~\bibnamefont {Fazio}}, \ and\ \bibinfo {author}
  {\bibfnamefont {P.}~\bibnamefont {Zoller}},\ }\bibfield  {title} {\enquote
  {\bibinfo {title} {Dynamical {{Phase Transitions}} and {{Instabilities}} in
  {{Open Atomic Many}}-{{Body Systems}}},}\ }\href {\doibase
  10.1103/PhysRevLett.105.015702} {\bibfield  {journal} {\bibinfo  {journal}
  {Phys. Rev. Lett.}\ }\textbf {\bibinfo {volume} {105}},\ \bibinfo {pages}
{015702} (\bibinfo {year} {2010})}\BibitemShut {NoStop}%
\bibitem [{\citenamefont {Tomadin}, \citenamefont {Diehl},\ and\ \citenamefont
{Zoller}(2011)}]{TomadinZoller2011}%
\BibitemOpen
\bibfield  {author} {\bibinfo {author} {\bibfnamefont {A.}~\bibnamefont
  {Tomadin}}, \bibinfo {author} {\bibfnamefont {S.}~\bibnamefont {Diehl}}, \
  and\ \bibinfo {author} {\bibfnamefont {P.}~\bibnamefont {Zoller}},\
  }\bibfield  {title} {\enquote {\bibinfo {title} {Nonequilibrium phase diagram
  of a driven and dissipative many-body system},}\ }\href {\doibase
  10.1103/PhysRevA.83.013611} {\bibfield  {journal} {\bibinfo  {journal} {Phys.
  Rev. A}\ }\textbf {\bibinfo {volume} {83}},\ \bibinfo {pages} {013611}
(\bibinfo {year} {2011})}\BibitemShut {NoStop}%
\bibitem [{\citenamefont {Yi}\ \emph {et~al.}(2012)\citenamefont {Yi},
  \citenamefont {Diehl}, \citenamefont {Daley},\ and\ \citenamefont
{Zoller}}]{YiZoller2012}%
\BibitemOpen
\bibfield  {author} {\bibinfo {author} {\bibfnamefont {W.}~\bibnamefont
  {Yi}}, \bibinfo {author} {\bibfnamefont {S.}~\bibnamefont {Diehl}}, \bibinfo
  {author} {\bibfnamefont {A.~J.}\ \bibnamefont {Daley}}, \ and\ \bibinfo
{author} {\bibfnamefont {P.}~\bibnamefont {Zoller}},\ }\bibfield  {title}
{\enquote {\bibinfo {title} {Driven-dissipative many-body pairing states for
  cold fermionic atoms in an optical lattice},}\ }\href {\doibase
  10.1088/1367-2630/14/5/055002} {\bibfield  {journal} {\bibinfo  {journal}
  {New J. Phys.}\ }\textbf {\bibinfo {volume} {14}},\ \bibinfo {pages} {055002}
(\bibinfo {year} {2012})}\BibitemShut {NoStop}%
\bibitem [{\citenamefont {Bardyn}\ \emph {et~al.}(2013)\citenamefont {Bardyn},
  \citenamefont {Baranov}, \citenamefont {Kraus}, \citenamefont {Rico},
  \citenamefont {{\textbackslash}.Imamo{\u g}lu}, \citenamefont {Zoller},\ and\
\citenamefont {Diehl}}]{BardynDiehl2013}%
\BibitemOpen
\bibfield  {author} {\bibinfo {author} {\bibfnamefont {C.-E.}\ \bibnamefont
  {Bardyn}}, \bibinfo {author} {\bibfnamefont {M.~A.}\ \bibnamefont {Baranov}},
  \bibinfo {author} {\bibfnamefont {C.~V.}\ \bibnamefont {Kraus}}, \bibinfo
  {author} {\bibfnamefont {E.}~\bibnamefont {Rico}}, \bibinfo {author}
  {\bibfnamefont {A.}~\bibnamefont {{\textbackslash}.Imamo{\u g}lu}}, \bibinfo
  {author} {\bibfnamefont {P.}~\bibnamefont {Zoller}}, \ and\ \bibinfo {author}
  {\bibfnamefont {S.}~\bibnamefont {Diehl}},\ }\bibfield  {title} {\enquote
  {\bibinfo {title} {Topology by dissipation},}\ }\href {\doibase
  10.1088/1367-2630/15/8/085001} {\bibfield  {journal} {\bibinfo  {journal}
  {New J. Phys.}\ }\textbf {\bibinfo {volume} {15}},\ \bibinfo {pages} {085001}
(\bibinfo {year} {2013})}\BibitemShut {NoStop}%
\bibitem [{\citenamefont {Lloyd}(1996)}]{Lloyd1996}%
\BibitemOpen
\bibfield  {author} {\bibinfo {author} {\bibfnamefont {S.}~\bibnamefont
    {Lloyd}},\ }\bibfield  {title} {\enquote {\bibinfo {title} {Universal
{{Quantum Simulators}}},}\ }\href {\doibase 10.1126/science.273.5278.1073}
{\bibfield  {journal} {\bibinfo  {journal} {Science}\ }\textbf {\bibinfo
  {volume} {273}},\ \bibinfo {pages} {1073--1078} (\bibinfo {year}
{1996})}\BibitemShut {NoStop}%
\bibitem [{\citenamefont {Yusipov}\ \emph {et~al.}(2017)\citenamefont
  {Yusipov}, \citenamefont {Laptyeva}, \citenamefont {Denisov},\ and\
\citenamefont {Ivanchenko}}]{YusipovIvanchenko2017}%
\BibitemOpen
\bibfield  {author} {\bibinfo {author} {\bibfnamefont {I.}~\bibnamefont
  {Yusipov}}, \bibinfo {author} {\bibfnamefont {T.}~\bibnamefont {Laptyeva}},
  \bibinfo {author} {\bibfnamefont {S.}~\bibnamefont {Denisov}}, \ and\
\bibinfo {author} {\bibfnamefont {M.}~\bibnamefont {Ivanchenko}},\ }\bibfield
{title} {\enquote {\bibinfo {title} {Localization in {{Open Quantum
    Systems}}},}\ }\href@noop {} {\bibfield  {journal} {\bibinfo  {journal}
  {Phys. Rev. Lett.}\ }\textbf {\bibinfo {volume} {118}},\ \bibinfo {pages}
{070402} (\bibinfo {year} {2017})}\BibitemShut {NoStop}%
\bibitem [{\citenamefont {Vershinina}\ \emph {et~al.}(2017)\citenamefont
  {Vershinina}, \citenamefont {Yusipov}, \citenamefont {Denisov}, \citenamefont
{Ivanchenko},\ and\ \citenamefont {Laptyeva}}]{VershininaLaptyeva2017}%
\BibitemOpen
\bibfield  {author} {\bibinfo {author} {\bibfnamefont {O.~S.}\ \bibnamefont
    {Vershinina}}, \bibinfo {author} {\bibfnamefont {I.~I.}\ \bibnamefont
  {Yusipov}}, \bibinfo {author} {\bibfnamefont {S.}~\bibnamefont {Denisov}},
  \bibinfo {author} {\bibfnamefont {M.~V.}\ \bibnamefont {Ivanchenko}}, \ and\
  \bibinfo {author} {\bibfnamefont {T.~V.}\ \bibnamefont {Laptyeva}},\
  }\bibfield  {title} {\enquote {\bibinfo {title} {Control of a single-particle
  localization in open quantum systems},}\ }\href {\doibase
  10.1209/0295-5075/119/56001} {\bibfield  {journal} {\bibinfo  {journal}
  {Europhys. Lett.}\ }\textbf {\bibinfo {volume} {119}},\ \bibinfo {pages}
{56001} (\bibinfo {year} {2017})}\BibitemShut {NoStop}%
\bibitem [{Note1()}]{Note1}%
\BibitemOpen
\bibinfo {note} {We assume no degeneracies}\BibitemShut {NoStop}%
\bibitem [{\citenamefont {Fan}\ \emph {et~al.}(2017)\citenamefont {Fan},
  \citenamefont {Zhang}, \citenamefont {Shen},\ and\ \citenamefont
{Zhai}}]{FanZhai2017}%
\BibitemOpen
\bibfield  {author} {\bibinfo {author} {\bibfnamefont {R.}~\bibnamefont
  {Fan}}, \bibinfo {author} {\bibfnamefont {P.}~\bibnamefont {Zhang}}, \bibinfo
  {author} {\bibfnamefont {H.}~\bibnamefont {Shen}}, \ and\ \bibinfo {author}
  {\bibfnamefont {H.}~\bibnamefont {Zhai}},\ }\bibfield  {title} {\enquote
  {\bibinfo {title} {Out-of-time-order correlation for many-body
  localization},}\ }\href {\doibase 10.1016/j.scib.2017.04.011} {\bibfield
  {journal} {\bibinfo  {journal} {Science Bulletin}\ }\textbf {\bibinfo
  {volume} {62}},\ \bibinfo {pages} {707--711} (\bibinfo {year}
{2017})}\BibitemShut {NoStop}%
\bibitem [{NSC()}]{NSCC}%
\BibitemOpen
\href@noop {} {}\bibinfo {howpublished} {https://www.nscc.sg/}\BibitemShut
{NoStop}%
\end{thebibliography}
\end{document}